\documentclass[review,number,sort&compress]{elsarticle}
\usepackage[utf8]{inputenc}
\usepackage{graphicx} 
\usepackage{caption} 
\usepackage{subcaption} 
\usepackage{enumerate}
\usepackage{multirow}
\usepackage{siunitx}
\usepackage{amsmath}
\usepackage{appendix}
\usepackage{lineno}
\usepackage{amssymb}

\newcommand\blfootnote[1]{%
  \begingroup
  \renewcommand\thefootnote{}\footnote{#1}%
  \addtocounter{footnote}{-1}%
  \endgroup
}

\newcommand{\event}{\itshape event\upshape}

\newcommand{\hit}{\itshape hit\upshape}
\newcommand{\hits}{\itshape hits\upshape}
\newcommand{\track}{\itshape track\upshape}
\newcommand{\tracks}{\itshape tracks\upshape}
\newcommand{\x}{$x$}
\newcommand{\y}{$y$}
\newcommand{\z}{$z$}
\newcommand{\micron}{\si{\micro \meter}}
\newcommand{\micros}{\si{\micro}s}

\journal{Nuclear Instruments and Methods A}

\begin{document}

\begin{frontmatter}

\title{Reducing DRIFT Backgrounds with a Submicron Aluminized-Mylar Cathode}
\author[well]{J.B.R.~Battat}
\author[shef]{E.~Daw}
\author[csu]{A.~Dorofeev}
\author[shef]{A.C.~Ezeribe}
\author[oxy]{J.R.~Fox}
\author[oxy]{J.-L.~Gauvreau}
\author[unm]{M.~Gold}
\author[oxy]{L.~Harmon}
\author[csu]{J.~Harton}
\author[unm]{R.~Lafler}
\author[oxy]{J.~Landers}
\author[unm]{R.J.~Lauer}
\author[unm]{E.R.~Lee}
\author[unm]{D.~Loomba}
\author[oxy]{A.~Lumnah}
\author[unm]{J.~Matthews}
\author[unm]{E.H.~Miller\corref{cor1}}
\ead{ehmiller@unm.edu, +016175714847}
\author[shef]{F.~Mouton}
\author[edin]{A.St.J.~Murphy}
\author[boul]{S.M.~Paling}
\author[unm]{N.~Phan}
\author[shef]{S.W.~Sadler}
\author[shef]{A.~Scarff}
\author[csu]{F.G.~Schuckman~II}
\author[oxy]{D.~Snowden-Ifft}
\author[shef]{N.J.C.~Spooner}
\author[shef]{D.~Walker}

\cortext[cor1]{Corresponding author}
\address[well]{Department of Physics, Wellesley College, Wellesley, MA 02481, USA}
\address[shef]{Department of Physics and Astronomy, University of Sheffield, Sheffield, S3 7RH, UK}
\address[csu]{Department of Physics, Colorado State University, Fort Collins, CO 80523, USA}
\address[oxy]{Department of Physics, Occidental College, Los Angeles, CA 90041, USA}
\address[unm]{Department of Physics and Astronomy, University of New Mexico, Albuquerque, NM 87131, USA}
\address[edin]{SUPA, School of Physics and Astronomy, University of Edinburgh, Edinburgh, EH9 3JZ, UK}
\address[boul]{STFC Boulby Underground Science Facility, Boulby Mine, Cleveland, TS13 4UZ, UK}


\begin{abstract}
\blfootnote{  }
Background events in the DRIFT-IId dark matter detector, mimicking potential WIMP signals, are predominantly caused by
alpha decays on the central cathode in which the alpha particle is completely or partially absorbed by the cathode material.  
We installed a 0.9 \micron~thick aluminized-mylar cathode as a way to reduce the probability of producing these backgrounds.  
We study three generations of cathode (wire, thin-film, and radiologically clean thin-film) with a focus on the ratio of background
events to alpha decays.  
Two independent methods of measuring the absolute alpha decay rate are used to ensure an accurate result, and agree to within $10\%$.  
Using alpha range spectroscopy, we measure the radiologically cleanest cathode version to have a contamination of $3.3\pm0.1$ ppt $^{234}$U and 
$73\pm2$ ppb $^{238}$U.  
This cathode reduces the probability of producing an RPR from an alpha decay by a factor of $70\pm20$ compared to the original stainless steel wire cathode.  
First results are presented from a texturized version of the cathode, intended to be even more transparent to alpha particles.  
These efforts, along with other background reduction measures, have resulted in a drop in the observed background rate from 500/day to 1/day.  
With the recent implementation of full-volume fiducialization, these remaining background events are identified, allowing for background-free operation.
\end{abstract}


\end{frontmatter}

\section{Introduction}
The properties of dark matter continue to be among the greatest outstanding mysteries in cosmology and particle physics.  The evidence for 
non-Baryonic dark matter is extensive \cite{Bertone2005}, and a Weakly Interacting Massive Particle (WIMP) is a
well-motivated candidate \cite{Jungman1996}.  
A convincing direct detection, which would 
both conclusively confirm the existence of WIMP dark matter and provide valuable information about its properties, proves to be elusive.  
Some recent \cite{CoGeNT2012} and older \cite{Bernabei2008} experimental results are suggestive of dark matter-like signals.  
These signals, despite being consistent with dark matter, have been called into question, demonstrating the need for a more convincing 
dark matter signature such as the sidereal modulation of the direction of incoming WIMP particles \cite{Spergel1988, Battat2009}.  
The Directional Recoil Identification From Tracks (DRIFT) dark matter experiment is the world's leading directional dark matter detector and is 
designed to provide an unambiguous detection of dark matter.  
DRIFT has demonstrated directionality down to 40 keVr \cite{Burgos2009, Turk2008} and set a spin-dependent limit (WIMP on proton) of 1.1 pb
for a 100 GeV/c$^2$ WIMP \cite{Battat2014}. 
The main challenge for DRIFT over the past 8 years has been radon progeny recoils (RPRs) from the central cathode, 
which have resulted in background rates as large as 500 events/day \cite{Burgos2007}.

Two major advances have eliminated these backgrounds, resulting 
in essentially background-free operation for DRIFT. 
The first 
advance has been the development of a thin film cathode that 
has contributed to a two-order-of-magnitude reduction in the background rate. 
The backgrounds and their reduction by the thin film cathode is the subject of this work.
The second 
was the discovery of a method for fiducialization along the drift direction \cite{Snowden2014}, which
enabled events near the cathode to be excluded as dark matter candidates. This method was implemented underground in the DRIFT-IId 
detector and demonstrated to work \cite{Miller2014a, Battat2014}. 

This paper begins with a description of the DRIFT detector (Section \ref{sec:drift}) and how alpha-decays at the cathode produce dark
matter backgrounds in DRIFT. 
This will be followed (Section \ref{sec:Analysis}) by a description of the analysis techniques based on alpha range spectroscopy
that identify both the isotopes and their location.  
This analysis was a critical tool in providing quantitative feedback on the
efficacy of the different versions of the thin film cathode in reducing the backgrounds. 
Section \ref{sec:ThinFilm} shows how the thin film cathode is expected to reduce the backgrounds in DRIFT-IId. 
These analysis tools were used to measure the radioactive contamination of the thin-film central cathode down to the ppt level 
(Section \ref{sec:Location}) and use this information to build a newer, cleaner version (Section \ref{sec:Clean}).  
Finally, we quantify improvements made by two versions of the thin film cathode, which has culminated in a background rate of $\approx1$ event/day.

\section{The DRIFT-IId Detector}
\label{sec:drift}
The DRIFT-IId Dark Matter detector is a 1~m$^3$ negative-ion time projection chamber (NITPC) which, while collecting data presented here, 
operated at a pressure of 40 Torr \cite{Alner2005}.
The bulk of the gas was 30 Torr of CS$_2$, an electronegative gas which captures ionized electrons, producing negative ions \cite{Snowden2000}.  
These negative ions drift $10^3$ times slower than electrons and with minimum (at the thermal limit) diffusion \cite{Martoff2000, Snowden2013}.  
The remaining 10 Torr of gas was CF$_4$, chosen for its high content of $^{19}$F providing spin-dependent sensitivity \cite{Ellis1991}.  

\begin{figure}[t] 
  \begin{subfigure}[t]{0.47\linewidth}
    \centering
    \includegraphics[width=0.85\linewidth]{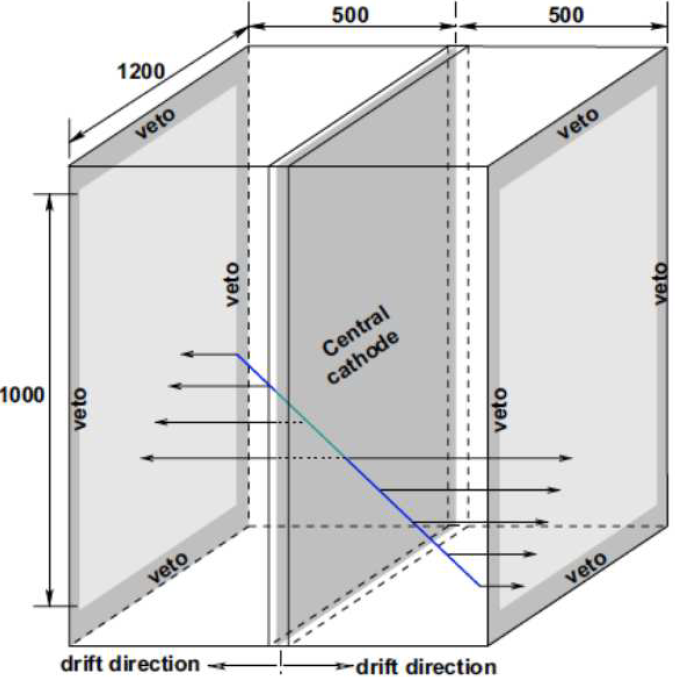} 
    \caption{Schematic of the DRIFT-IId detector, showing a cathode-crossing alpha track.  Lengths are in mm.  Image reproduced from \cite{Burgos2008}.  } 
    \label{fig:DIISchematic} 
  \end{subfigure}
  \hfill
  \begin{subfigure}[t]{0.47\linewidth}
    \centering
    \includegraphics[width=1.1\linewidth]{DIIphoto.png} 
    \caption{Photograph of the DRIFT-IId detector removed from its vacuum vessel.  } 
    \label{fig:DIIPhoto} 
  \end{subfigure} 
  \caption{ The DRIFT-IId detector.}
  \label{fig:DII} 
\end{figure}

The DRIFT-IId detector (Figure \ref{fig:DII}) contains two 50-cm deep detection volumes that share a single central cathode plane.  
During the data collection runs analyzed in this document, the cathode plane was at -30242 V which, 
together with a wire field cage, defines a uniform drift field of E=550 V/cm in each volume.
Each volume was
terminated by a 1~m$^2$ Multi-Wire Proportional Chamber (MWPC) which is comprised of three
parallel planes separated by 1~cm.  The middle anode plane was originally built from $20$~\si{\micro \meter} stainless steel wires at ground potential.  
The two outer grid planes use $100$~\si{\micro \meter} wires oriented perpendicular to the anodes and held at -2731 V.  
This voltage difference provides gas amplification with a gain of $\sim1000$.  

\begin{figure}[h]
 \centering
 \includegraphics[width=0.85\textwidth]{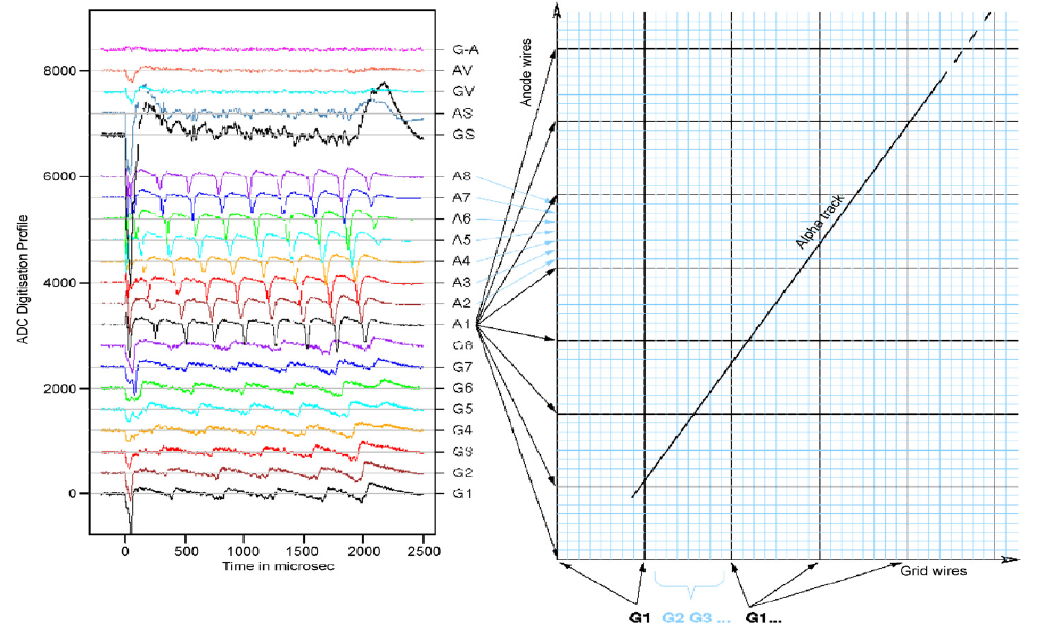}
 \caption{ The DRIFT detector groups all of its 448 anode wires (horizontal on right) periodically into 8 readout lines (A1-A8 on left), and
does the same for the perpendicular grid wires (vertical on right; G1-G8 on left).  With a wire pitch of 2 mm, this introduces a periodicity in space of 16 mm which
is much larger than the length of a typical nuclear recoil.  Alpha particle tracks, however, can be hundreds of mm in length and typically hit each of the 8 channels more than
once.  Above are the data from a real alpha track (left) compared with a diagram showing the projected alpha track (right).  Note that a signal pulse may overlap
the electronic overshoot from a previous pulse, confusing the charge (but not length) measurement of such long tracks.  }
 \label{fig:AlphaProjection}
\end{figure}

Each of these planes has 552 wires with a pitch of 2~mm.  The outermost 52 (41) wires of the anode (grid) are used to identify and veto events entering the 
fiducial volume of the detector from the outside.  The remaining 448 (459) wires in each plane, spanning a 896~mm (918~mm) fiducial length, are grouped into 8 channels such that 
every eighth wire is read out by the same channel; this introduces a periodicity of 16 mm in the readouts.  This does not affect the WIMP search as the low energy 
nuclear recoils of interest have tracks that are typically less than 5~mm long.  This periodicity can be seen in longer tracks, such as those from alpha particles
or protons, as seen in Figure \ref{fig:AlphaProjection}.  The 8 anode channels measure the track along the $x$ axis while the perpendicular
grid channels measure the $y$ extent of a track.  The $z$ component of the length is measured by the transit of charge into the MWPC at a known drift speed of $59.37\pm0.15$ m/s.  
The digitization rate of 1~MHz and fast electronics correspond to a sub-mm spatial resolution along the \z~ axis, 
so the measurement of the track along this axis is the most precise and accurate.  

The detector is located in the Boulby Underground Laboratory, at 2805 m.w.e., to shield from cosmogenics \cite{Boulby2012}.  
It is further shielded from rock neutrons by 
polypropylene pellets providing at least 35~g~cm$^{-2}$ of hydrogenous shielding.  
This is expected to reduce backgrounds from rock neutrons to less than 1/year \cite{Burgos2007}.

\subsection{DRIFT Backgrounds and the Thin-Film Solution}
\label{sec:RPRs}

The first studies of background events in DRIFT-II observed a prohibitively high rate of WIMP-like backgrounds; around 500/day \cite{Burgos2007}.  
These have been attributed primarily to Radon Progeny Recoils (RPRs) produced at the 20~\si{\micro \meter} stainless steel wires of the cathode used at the time.  
Production of an RPR often begins with the emanation of $^{222}$Rn inside the vacuum vessel. 
The $^{222}$Rn atom diffuses into the fiducial volume and decays, emitting a 5.49 MeV alpha particle and a $^{218}$Po atom which is typically positively charged 
(the charged fraction and its measurement is described in Section \ref{sec:neutral}).
While this initial alpha particle track is easily identified, it is the charged $^{218}$Po that has the potential to initiate the RPR backgrounds.  
After the $^{218}$Po drifts to and electrodeposits on the central cathode, it alpha-decays with a half-life of 3.05 minutes, 
emitting a 6.00 MeV alpha and a 112 keV $^{214}$Pb atom.
Alpha particles of this energy have a range of 12.4~\si{\micro \meter} in stainless steel and, for isotropic decays, 
will range out in the 20~\si{\micro \meter} wire $34\%$ of the time.  
The lead atom, meanwhile, is ejected into the gas and produces an RPR background.  

\begin{figure}[t]
 \centering
 \includegraphics[width=0.75\textwidth]{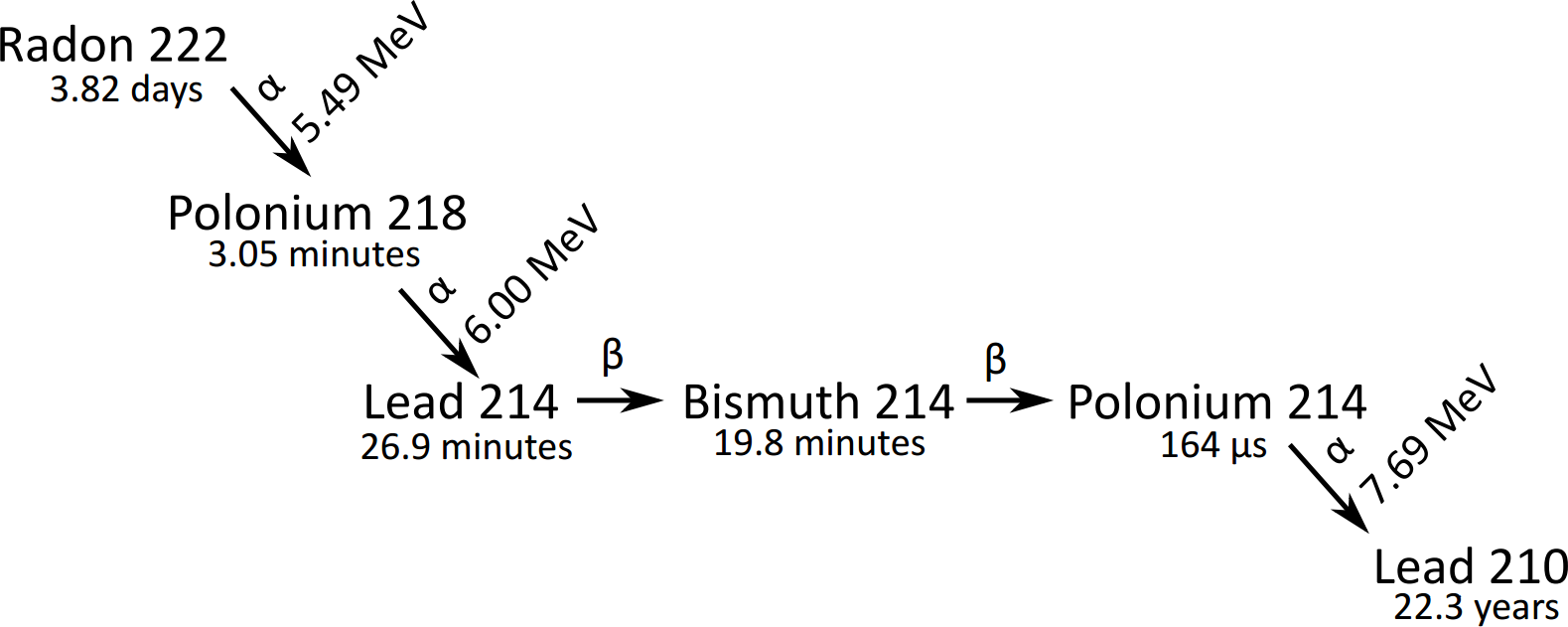}
 \caption{ Decay sequence from $^{222}$Rn. The decays shown here are preferred with branching fractions $>99\%$.   } 
 \label{fig:DecayChain}
\end{figure}

\begin{figure}[t]
 \centering
 \includegraphics[width=0.65\textwidth]{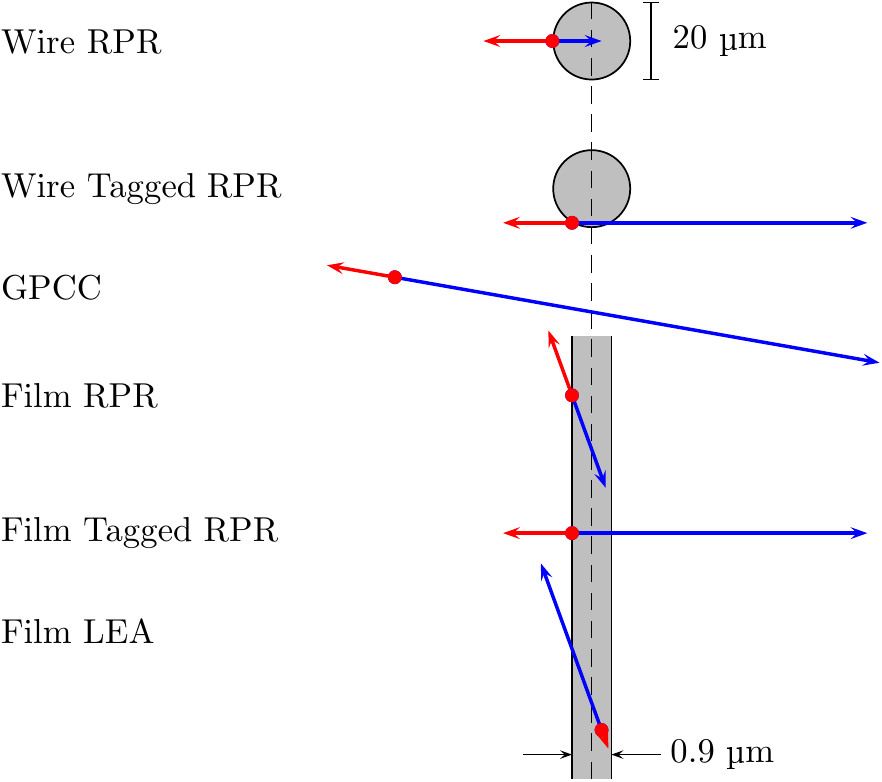}
 \caption{ Schematic of various alpha decay event types which are discussed in the text.  
 Shaded in gray are cross sections of the central cathode, with 20~\micron~wires on top and 0.9~\micron~aluminized mylar on the bottom; 
 note that the steel is thicker and provides more stopping power than the thin-film.  
 Both the alpha particle track (blue) and the recoiling daughter track (red) are presented.  
 GPCCs, alphas which cross the cathode plane, occur with both cathode geometries.  
 Radon progeny recoils (RPRs) are the dominant background for the wire cathode, while both these and low-energy alphas (LEAs) 
 contribute to backgrounds with the thin-film cathode.
 Image is not to scale.  }
 \label{fig:Backgrounds}
\end{figure}

Two more RPRs are possible from the same $^{222}$Rn progenitor (see Figure \ref{fig:DecayChain}).  
Over the course of an hour $^{214}$Pb undergoes two $\beta$ decays to become
$^{214}$Po, which then alpha-decays into $^{210}$Pb and produces a 7.69~MeV alpha.  
This Pb atom, with a half-life of 22.3 years, may, over the life of the detector, 
decay by two $\beta$ emissions into $^{210}$Po which produces one final alpha decay.  
The resultant stable daughter of this chain is $^{206}$Pb. 
Thus, from the initial $^{222}$Rn, the $^{214}$Pb, $^{210}$Pb, and $^{206}$Pb all have potential to produce an RPR background from the central cathode of DRIFT-IId. 
While similar background-producing decay chains are possible from other radon isotopes (such as $^{220}$Rn), the majority of the alpha-producing radon
progeny decays at the cathode are daughters of $^{222}$Rn \cite{Burgos2008}, as will be shown in Section \ref{sec:Analysis}.\footnote{In 2007, 
a large population of $^{210}$Po (with a 22.2 year half-life) on the surface of the central cathode wires contributed significantly to the background
rate in DRIFT-IId.  This population is not present on more recent cathodes, and the long half-life of the decay limits the impact of its buildup 
from radon decays in the volume \cite{Sadler2014c}.  }

The RPR hypothesis is now supported by a wealth of data.  
These background events have been observed to suffer more longitudinal diffusion than neutron-induced nuclear recoils in the bulk of the gas, 
 which suggests that these events originate near the cathode (the maximal drift distance) \cite{Daw2010}.  
Long-term background studies have revealed a correlation between the rates of $^{222}$Rn decays and background events in DRIFT-IId \cite{Sadler2014c}.
A nitric acid etch of the wire central cathode, to remove radon progeny and any other alpha decay contaminants from the surface, reduced the background
 rate \cite{Daw2010}, while a similar etch of the MWPCs had only a minor effect on the background rate \cite{Turk2008, Sadler2014c}.  
Finally, the discovery of minority carriers, which resulted in the ability to fiducialize events along the drift direction, 
 has definitively shown that the RPRs occur at the cathode \cite{Miller2014a, Battat2014}.

As shown in Figure \ref{fig:Backgrounds}, an RPR event at the cathode requires both that the alpha is fully absorbed in the cathode and that the Pb
recoil enters and deposits a detectable amount of energy in the gas.  
Due to the short range ($\mathcal{O}(100$~\AA)) of the Pb recoil in the cathode \cite{SRIM}, the RPR must originate at or very near to the surface of the cathode. 
Given the double-sided chamber employed in DRIFT-IId, there are other event topologies possible due to alpha decays at or in the central cathode 
(see Figure \ref{fig:Backgrounds}).  
These will be discussed later (Sections \ref{sec:Analysis} and \ref{sec:LEA}), but it is important to mention the tagged RPR which is present in all DRIFT cathodes.
This is identical to a regular RPR except that the alpha particle is also observed on the opposite side of the detector.  
The observation of the alpha particle track in coincidence with the RPR allows us to tag the RPR and exclude it as a WIMP candidate.
Studying this and other classes of identifiable alpha decays at the central cathode provides useful information as described in the following sections.

The DRIFT collaboration has devised various strategies to  mitigate and eliminate these background events. 
The first was to identify and replace radon-emanating materials with less radioactive ones. 
In addition, we performed a nitric etch of the wire cathode, which removed long-lived surface contaminants introduced in the manufacturing as well as
accumulated $^{210}$Po \cite{Turk2008}.  
This, together with the replacement of radon-emanating materials, reduced the background rate from 500/day to 130/day in 2009 \cite{Daw2010}.  
Efforts to reduce the rate of radon emanation in DRIFT have continued, culminating in a factor of 10 reduction since 2005 \cite{Sadler2014b}.  

The strategy described in this work is the replacement of the wire central 
cathode with a 0.9~\si{\micro \meter} aluminized thin film that is nearly transparent to the alphas produced in the radon progeny decays. 
This greatly reduces the chance that an RPR at the cathode will go untagged by the coincident alpha produced in the decay. 
This cathode has provided a further factor of 30 reduction in the background rate, down to $\approx$1/day. 
With the z-fiducialization described above, the remaining RPRs are now identified and removed from the dark matter data, 
enabling DRIFT to operate with zero backgrounds while preserving a good nuclear recoil efficiency.

As we  describe results from the various generations of cathodes it is useful to define a figure of merit for the cathode's efficacy 
to reject alpha-decay backgrounds. Our definition  is the ratio of observed background events at the cathode that are classified as 
nuclear recoils to the total number of alpha decays occurring on or in the cathode:

\begin{equation}
 \text{Figure of Merit} = \frac{\text{Measured Backgrounds}}{\text{All Alphas from Cathode}}
 \label{eq:FoM}
\end{equation}

As more RPRs are tagged with alphas, this quantity will decrease indicating a reduction in backgrounds due to an improved cathode. 
This definition is insensitive to differences in the RPR rate due to radon emanation changes in the detector that occurred over the same 
time-scale as the cathode R\&D. The numerator of this figure of merit, the measured backgrounds, is obtained from an independent nuclear recoil 
analysis similar to that presented in \cite{Daw2010}. The results are in agreement, with the exception that a higher threshold is employed here 
to focus on backgrounds due to alpha decays.  
The region of interest used here 
is 1000-4000 NIPS (Number of Ion Pairs), calibrated against the 237 NIPS produced by a 5.9 keV X-ray from a $^{55}$Fe source \cite{Pushkin2009}. 

To derive the total alpha decay rate for the denominator in the figure of merit, a knowledge of the detection efficiency for alphas in DRIFT is 
required to convert the observed alpha counts to an absolute number of decays. 
This is a new analysis that is central to this work, and is described in the next section. 
There we also show how DRIFT's 3D alpha range measurements can be used to do spectroscopy that gives detailed information on 
the specific isotopes that are responsible for the backgrounds and their location.

\section{Alpha Range Spectroscopy}
\label{sec:Analysis}

Over the years, alphas have provided an excellent tool to study backgrounds in DRIFT \cite{Burgos2008}.  
Many radioactive contaminant decay chains, such as the $^{222}$Rn chain, provide identifiable alphas.  
Furthermore, as described above, DRIFT's backgrounds come directly from alpha decays occurring on or in the central cathode. 
The basis of the alpha analysis to follow is the identification and measurement of alpha tracks in the DRIFT-IId detector. 
Alphas are used for spectroscopy based on their range rather than their energy. 
Alpha particle energy is difficult to measure with DRIFT's readout scheme, which has been optimized for much shorter and less energetic nuclear recoil events.

For each \event~the eight anode lines on both sides of the detector are scanned for voltage excursions from ionization. 
A \hit~is the signal on an anode wire beginning when the signal last rose above 0 before reaching $3\sigma_n$ for at least $3\micros$~and lasting
until it once more crosses the baseline.  The baseline gaussian noise, $\sigma_n$, is calculated independently for each of the 32 signal and 4 veto channels.  

Within the detector volume, ionization resulting from an alpha particle track can reach up to 56 cm long and 
is typically much longer than the 2 mm anode wire  separation.  
To compile the relevant \hits~from a single ionization event within DRIFT, we collect all \hits~on one side which overlap in time. 
This collection is a \track~and contains all of the data from one side of the detector corresponding to a single ionization event.  

These \tracks~are analyzed to classify them into well-understood categories including nuclear recoils, electron recoils, sparks and, 
most relevant to this work, alphas.  Any \track~long enough to produce 10 consecutive \hits~ (2 cm in \x),  
which does not reach a threshold of $5.5$~mV on either veto channel, and which does not cross the MWPC is classified as a contained alpha track.  
Events crossing an MWPC are identified by a sharp rise time as ionization is produced in the high-field region of the MWPC.  
If a \track~is otherwise contained
but does cross the central cathode, appearing as an alpha \track~on each MWPC, then it is called a Gold-Plated Cathode Crosser (GPCC) as shown in 
Figure \ref{fig:Backgrounds}.

The length components along all three axes are measured for each alpha track.  
The \z~length, along the drift direction, is calculated as $\Delta z = v_d * \Delta t$, 
where $\Delta t$ is the time difference between the start of the first \hit~and the end of the last \hit~in the \track. 
The drift speed, $v_d=59.37\pm0.15$ m/s, is measured in-situ by the timing of alpha particles from $^{214}$Po decays at the 
central cathode which cross an MWPC and are therefore known to travel exactly 500 mm in $z$.  

The \x~length, perpendicular to the anode wires, is determined by multiplying the number of \hits~on the anode by the wire spacing, 2 mm.  In a similar
fashion the \y~length, which is along the axis perpendicular to the grid wires, is measured by counting pulses on each wire and multiplying this sum
by the wire spacing, which is again 2 mm.  Here, a pulse is defined as a peak in the waveform extending forward and backward until the signal falls off by
$8\sigma_n$ ($\sigma_n$, discussed above, is the gaussian noise on the channel).  $8\sigma_n$ is used here to avoid induced signals from neighboring wires.  

\begin{figure}[t]
  \centering
  \begin{minipage}[t]{0.475\textwidth}
    \centering
    \includegraphics[width=\textwidth]{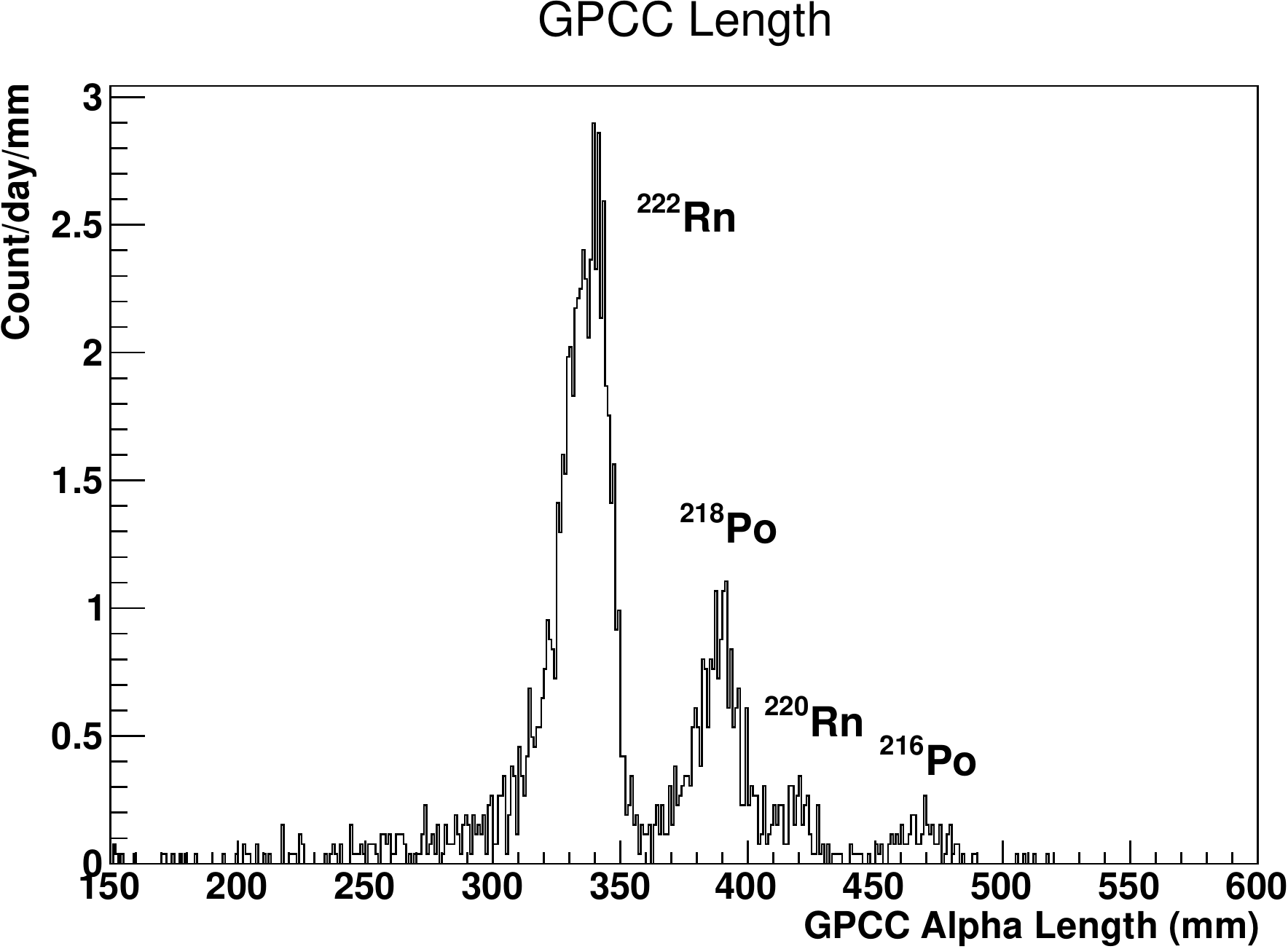}
    \caption{ Distribution of measured lengths of GPCC alpha particle tracks from the first thin-film runs.} 
    \label{fig:GPCC}
  \end{minipage}%
  \begin{minipage}[t]{0.05\textwidth}
  \hspace{\textwidth}
  \end{minipage}%
  \begin{minipage}[t]{0.475\textwidth}
    \centering
    \includegraphics[width=\textwidth]{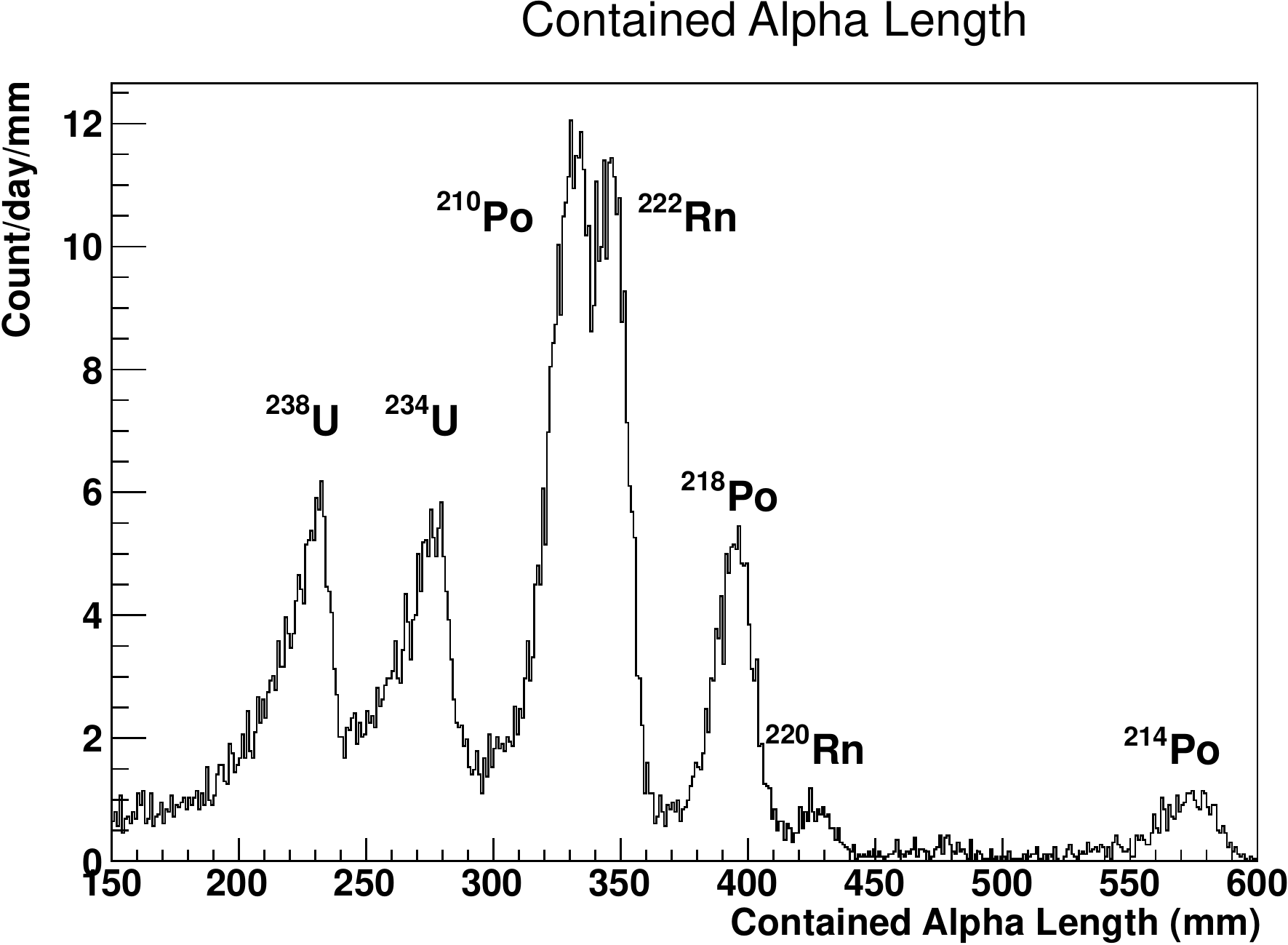}
    \captionof{figure}{ Distribution of measured lengths of contained (non-cathode-crossing) alpha particle tracks from the first thin-film runs.  }
    \label{fig:Cont}
  \end{minipage}
\end{figure}

Using this measure of the alpha length along all three axes, we histogram the full 3-d lengths of both contained and GPCC alphas within the detector 
(see Figures \ref{fig:GPCC}-\ref{fig:Cont}).  The spectrum of peaks in the histogram can be identified as decays from particular isotopes using the 
mean length of the alpha particle's range, obtained from SRIM \cite{SRIM}.  
This technique can measure the 3D range of alpha particle \tracks~in the detector with $5\%$ resolution.  
Due to the overlapping nature of many of the populations, the number of observed decays is calculated by integrating over
the best fit Crystal Ball function \cite{CrystalBall1, CrystalBall2, CrystalBall3}.\footnote{The 
long tail of the Crystal Ball function fits the range distribution from \tracks~which are partially absorbed by the thin-film cathode.}

The GPCC events shown in Figure \ref{fig:GPCC} are due to decays occurring in the gas and show that the radon diffusing into the 
fiducial volume of DRIFT-IId is dominated by $^{222}$Rn, with only a small amount of $^{220}$Rn. 
Conversely, the contained alpha distribution shows contributions from alpha decays occurring in the gas as well as at the cathode and MWPCs. 
The appearance of the uranium and lead isotopes points to contamination at the latter sites. 
We will show how to isolate the contributions of each population after discussing the alpha efficiency analysis in the next section.

\subsection{Alpha Measurement Efficiencies}
\label{sec:efficiency}

There are two major sources of inefficiency for alpha detection. 
First, the entire length of the alpha track must be contained in the fiducial volume of the detector. 
This is derived entirely from the geometry of the detector and the particle track and is called the Geometric Acceptance (GA).  
Second, the alpha must be correctly identified and classified as an alpha by the analysis.  This is affected by features of the detector's data
acquisition and the subsequent analysis and is called the Analysis Efficiency (AE).  
Both are inherently geometrical, with the first being due to 
the alpha track crossing a detector boundary, whereas the second arises due to different orientations of the alpha track with respect to the detector's 
principle axes. In the following we describe these in greater detail.  

For brevity, only the calculations for $^{222}$Rn GPCC alphas are presented, 
although efficiencies for other categories of alpha track are presented in Table \ref{tab:Efficiencies}.

\subsubsection{Geometric Acceptance}
\label{sec:ContFrac}

The GA is the probability that an alpha particle produced by a decay in the fiducial volume remains contained within it.  
The fraction of alpha tracks that are contained is determined analytically and checked against a simple geometric simulation.  
In the simulation, the origin of $10^7$ alpha particles are chosen randomly within the fiducial volume of the detector.  
They are each propagated in an isotropically distributed random direction, with a distance given by the average length for that isotope as 
measured by the range spectra (Figures \ref{fig:GPCC}-\ref{fig:Cont}).
If the end point is still within the fiducial volume and the track crossed the central cathode, it is classified a GPCC.  
The fraction of 348~mm alpha tracks from $^{222}$Rn decay which produce GPCCs that are fully contained is $0.121$.
This is the first of the two geometrical efficiencies discussed above.  

\subsubsection{Analysis Efficiency}
\label{sec:angles}

When an alpha track is oriented parallel to either the anode or grid wires, the individual periodic (see Section \ref{sec:drift}) \hits~on a 
single channel overlap and become difficult to distinguish.  Likewise, tracks oriented parallel to the drift direction (\z) have \hits~that are 
very long in time and are skewed by the electronics which have been designed for short nuclear recoils.  Furthermore, such a \track~may not 
produce \hits~on all eight channels, as required for it to be classified as an alpha track. Both of these effects make it difficult or impossible 
to identify these as alpha tracks and introduces an angle-dependent loss of efficiency. This is further divided into the efficiency along $\theta$, 
the zenith angle, and $\phi$, the azimuthal angle of the \track.  

\begin{figure}[t]
  \centering
  \begin{minipage}[t]{0.475\textwidth}
    \centering
    \includegraphics[width=\textwidth]{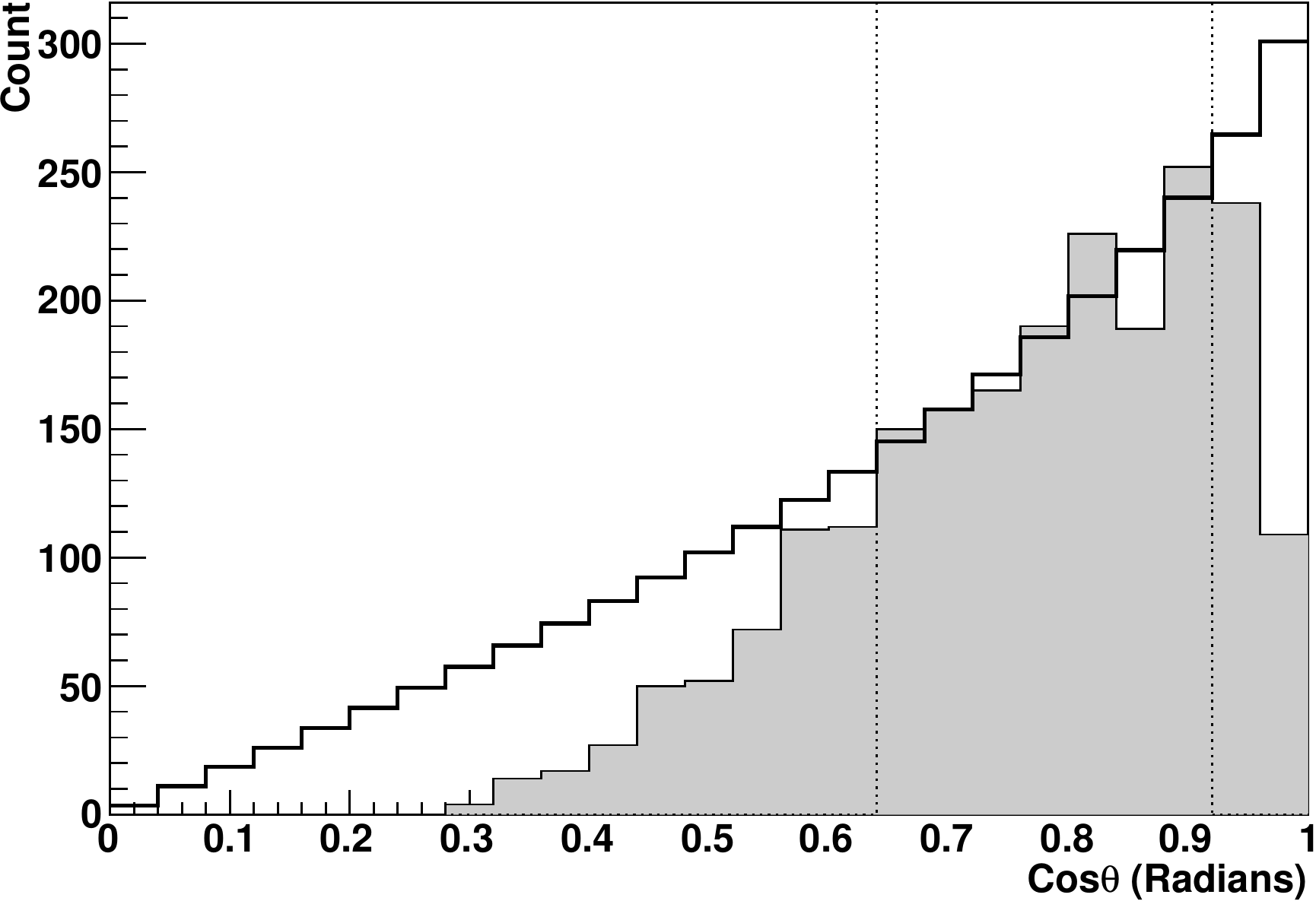}
      \captionof{figure}{ The theoretical distribution of zenith angles ($\theta$) of contained 348~mm-long cathode-crossing alphas is shown as a thick line;
      the same distribution obtained from DRIFT-IId data is shown as shaded gray.  The model distribution is scaled to match the distribution from data along 
      the 7 bins within the vertical dotted lines.  The final efficiency for $^{222}$Rn GPCC along $\theta$ is $0.733\pm0.011$  }
    \label{fig:Rn222Theta}
  \end{minipage}%
  \begin{minipage}[t]{0.05\textwidth}
  \hspace{\textwidth}
  \end{minipage}%
  \begin{minipage}[t]{0.475\textwidth}
    \centering
    \includegraphics[width=\textwidth]{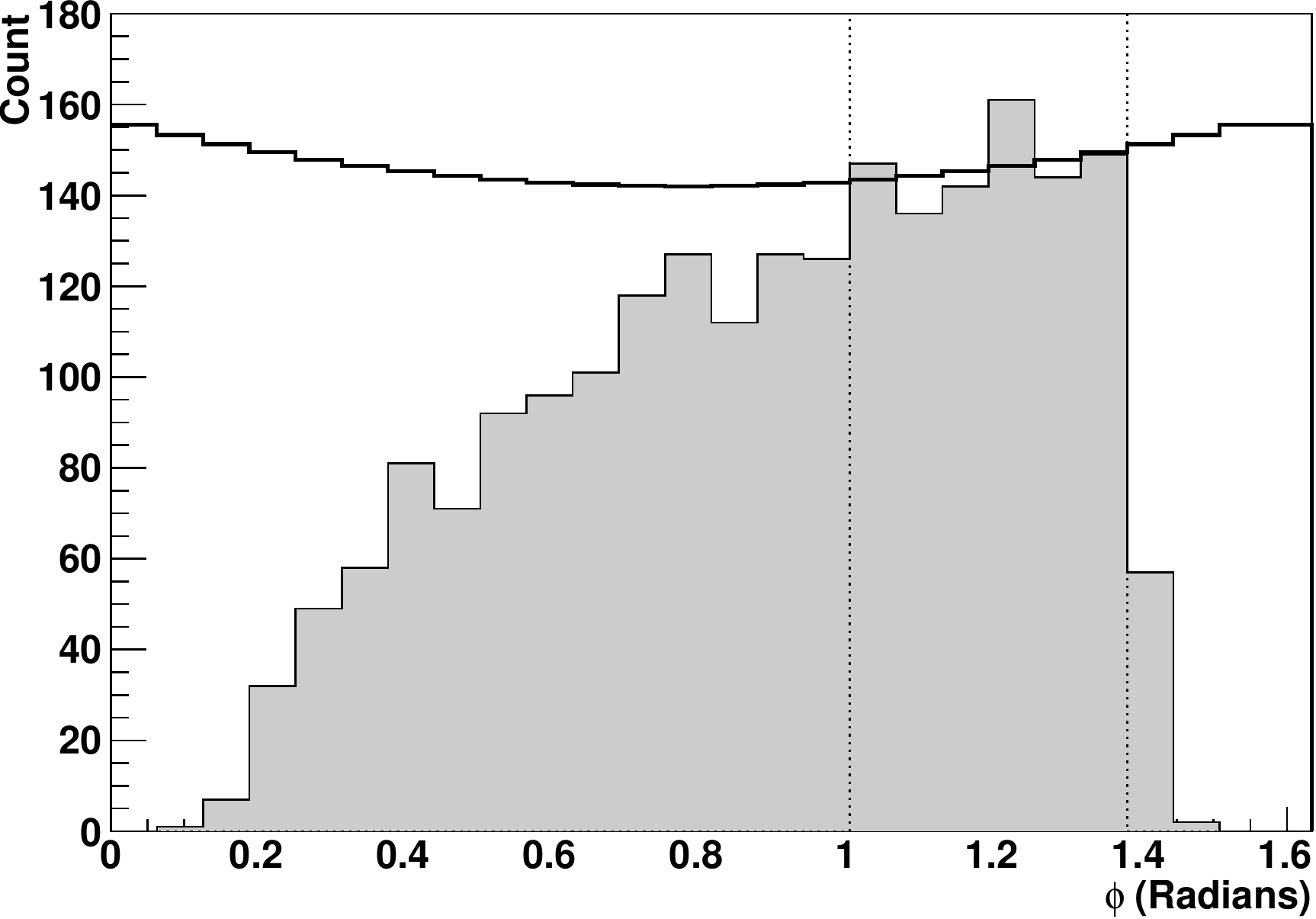}
      \captionof{figure}{ The theoretical distribution of azimuthal angles ($\phi$) of contained 348~mm-long cathode-crossing alphas is shown as a thick line;
      the same distribution obtained from DRIFT-IId data is shown as shaded gray.  The model distribution is scaled to match the distribution from data along 
      the 5 bins within the vertical dotted lines.  The final efficiency for $^{222}$Rn GPCC along $\phi$ is $0.582\pm0.014$.  }
    \label{fig:Rn222Phi}
  \end{minipage}
\end{figure}

These efficiencies are determined analytically, by calculating the distribution of angles for $^{222}$Rn GPCC alpha particles and comparing with the 
observed distributions (see Figs \ref{fig:Rn222Theta} - \ref{fig:Rn222Phi}).  For each distribution, a range of angles is chosen over which the 
identification efficiency is assumed to be 
$100\%$.\footnote{This assumption is verified by the agreement with the efficiency calculated using an independent method in Section \ref{sec:TCorr}}
The measured distribution is scaled to match the calculation and the overall efficiency over that dimension
is taken to be the ratio of areas of the distributions.  
The efficiency for detecting GPCC alphas from$^{222}$Rn decays is $0.733\pm0.011$ over $\theta$ and $0.582\pm0.014$ over $\phi$. 
Only statistical uncertainties are quoted here; systematic uncertainties will be estimated in Section \ref{sec:TCorr}.

The total AE, assuming the two angular efficiencies are uncorrelated, is the product of the efficiencies along these two orthogonal angles, $0.427\pm0.012$.  
This assumption introduces a systematic error of $\leq1\%$ into the calculation, 
estimated by running a simulation with input angles that are tailored to match measured distributions rather than isotropic. 
The total identification efficiency for $^{222}$Rn GPCCs is obtained by taking the product of the AE (0.472) and the
GA (0.121), yielding $0.0517\pm0.0015$.  
This efficiency is the probability for the production and subsequent identification of a $^{222}$Rn GPCC.   
In order to find the absolute number of radon decays in DRIFT's fiducial volume, one would take the observed number of GPCCs and divide by this efficiency.

\begin{table}[h]
\centering
 \begin{tabular}{l|r|rrr|r}
   Category & Length & Geom. & $\theta$ & $\phi$ & Total \\
   \hline
   $^{222}$Rn GPCC & 348 & 0.121 & 0.733& 0.582 & $0.052 \pm 0.001 $\\
   \hline
   $^{218}$Po GPCC & 396 & 0.130 & 0.772 & 0.618 & $0.062 \pm 0.005 $\\
   $^{218}$Po Cath & 396 & 0.599 & 0.597 & 0.700 & $0.250 \pm 0.008 $\\
   \hline
   $^{238}$U Cath & 234 & 0.753 & 0.630 & 0.792 & $0.376 \pm 0.002 $\\
   \hline
   $^{234}$U Cath & 279 & 0.709 & 0.566 & 0.731 & $0.293 \pm 0.009 $\\
 \end{tabular}
 \caption{ Summary of relevant efficiencies.  
The categories include various isotopes as either cathode-crossing (GPCC) or as originating from the central cathode (Cath).  
The lengths are measured in mm.  
Listed are the GA (Section \ref{sec:ContFrac}) and the two angle-dependent AEs (Section \ref{sec:angles}).  
The product of these three provides the total efficiency for detecting and identifying these alpha particle tracks, with statistical 
 uncertainties provided.  
All efficiencies also carry a systematic uncertainty of 10\%. }
 \label{tab:Efficiencies}
 
\end{table}

\subsubsection{Time Correlation}
\label{sec:TCorr}
The method of deriving the alpha detection efficiency described above relies on the critical assumption that there are regions of the $\theta$ and $\phi$ 
parameter space where the efficiency is $100\%$ (see Figures \ref{fig:Rn222Theta} and \ref{fig:Rn222Phi}). 
To test this, we have developed a second independent method that 
is based on examining correlations in time between consecutive alpha decays in the $^{222}$Rn decay chain.  
For example, $^{222}$Rn will alpha-decay, producing $^{218}$Po and a 5.49 MeV alpha.  
This polonium atom decays with a half-life of 3.05 minutes, producing a second alpha particle.  
Because these alpha particles have different lengths and may follow different spatial distributions, 
they are assumed to have different probabilities for detection, $\eta_{Rn}$ and $\eta_{Po}$, respectively.  
Due to the nature of the decay chain and the short half-life relative to the gas flow rate (usually 1 volume/day), 
we assume that the same absolute number $N$ of each decay occurs within the fiducial volume.  

The number of observed decays is the product of the total number of decays $N$ and the efficiency of observing that particular category of alpha. 
The probability of observing the alphas from both the decays of $^{222}$Rn and its daughter, $^{218}$Po, is then $\eta_{Rn}\eta_{Po}$.  
The efficiency of detecting one particular category (e.g. $^{222}$Rn) is equal to the number of parent-daughter 
pairs ($N_{pairs}$) observed divided by the total number ($N_{Po}$) of the other category observed (e.g. $^{218}$Po):
  
\begin{equation}
\label{eq:effi}
 \eta_{Rn} = \frac{N\eta_{Rn}\eta_{Po}}{N\eta_{Po}} = \frac{N_{pairs}}{N_{Po}}
\end{equation}

\begin{figure}[ht]
 \centering
 \includegraphics[width=0.55\textwidth]{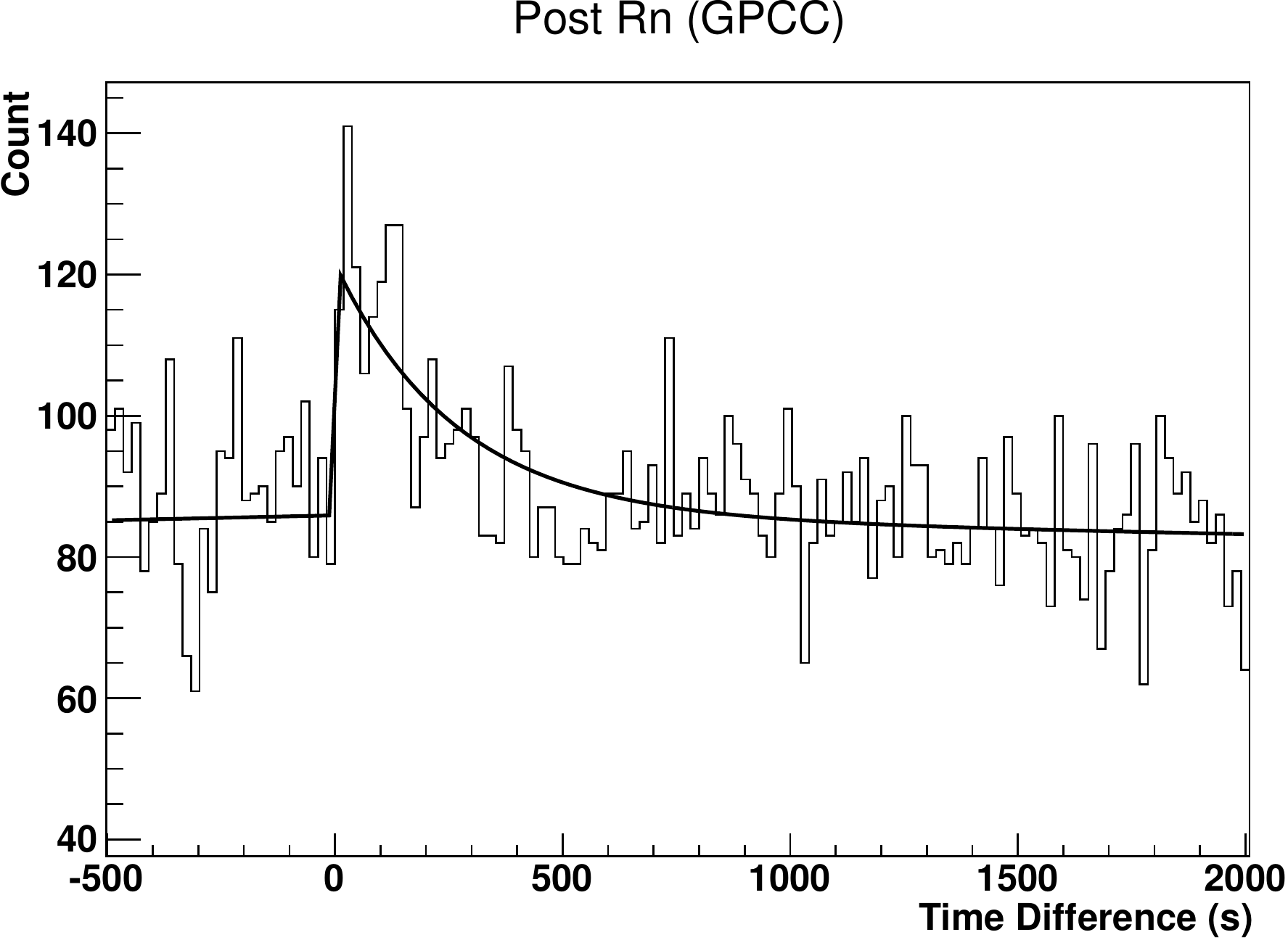}
 \caption{ A histogram of the timing difference, $t_{Po}-t_{Rn}$, between every possible pairing of $^{222}$Rn GPCC and $^{218}$Po.  When both tracks derive
from the decays of a single atom, the timing follows an exponential decay.  When the tracks derive from decays of different atoms, the timing
is uncorrelated and exhibits a nearly flat background.  The black curve is the best fit of the sum of an exponential decay and a background model.   }
 \label{fig:DecayCorr}
  \end{figure}

\ref{ap:TCorr} contains a more rigorous derivation.  
The number of pairs ($N_{pairs}$) can be measured through timing correlations.  
In this example we take timing differences between all possible pairs of $^{222}$Rn and $^{218}$Po decays. 
Only the true parent-daughter pairs give the characteristic exponential timing correlation 
with the 3.05 minute half-life of Po-218; the rest should be uncorrelated.
Figure \ref{fig:DecayCorr} 
shows a histogram in the timing difference of all observed $^{222}$Rn and $^{218}$Po pairs.  
This distribution is fit to an exponential decay starting at $t=0$ with a second-order polynomial background, symmetric around zero, 
to model the uncorrelated decays.  
Here the exponential fit integrates to 505 identified event pairs above the background.  
Compared with the 8830 $^{218}$Po alphas observed, this corresponds to a $0.058\pm0.003$ identification efficiency for $^{222}$Rn.  
However, in order to compare this efficiency with that obtained in Section \ref{sec:efficiency}, it is necessary to account for the potential difference in
spatial distributions of decays, described above.

\begin{figure}[t]
 \centering
 \includegraphics[width=0.7\textwidth]{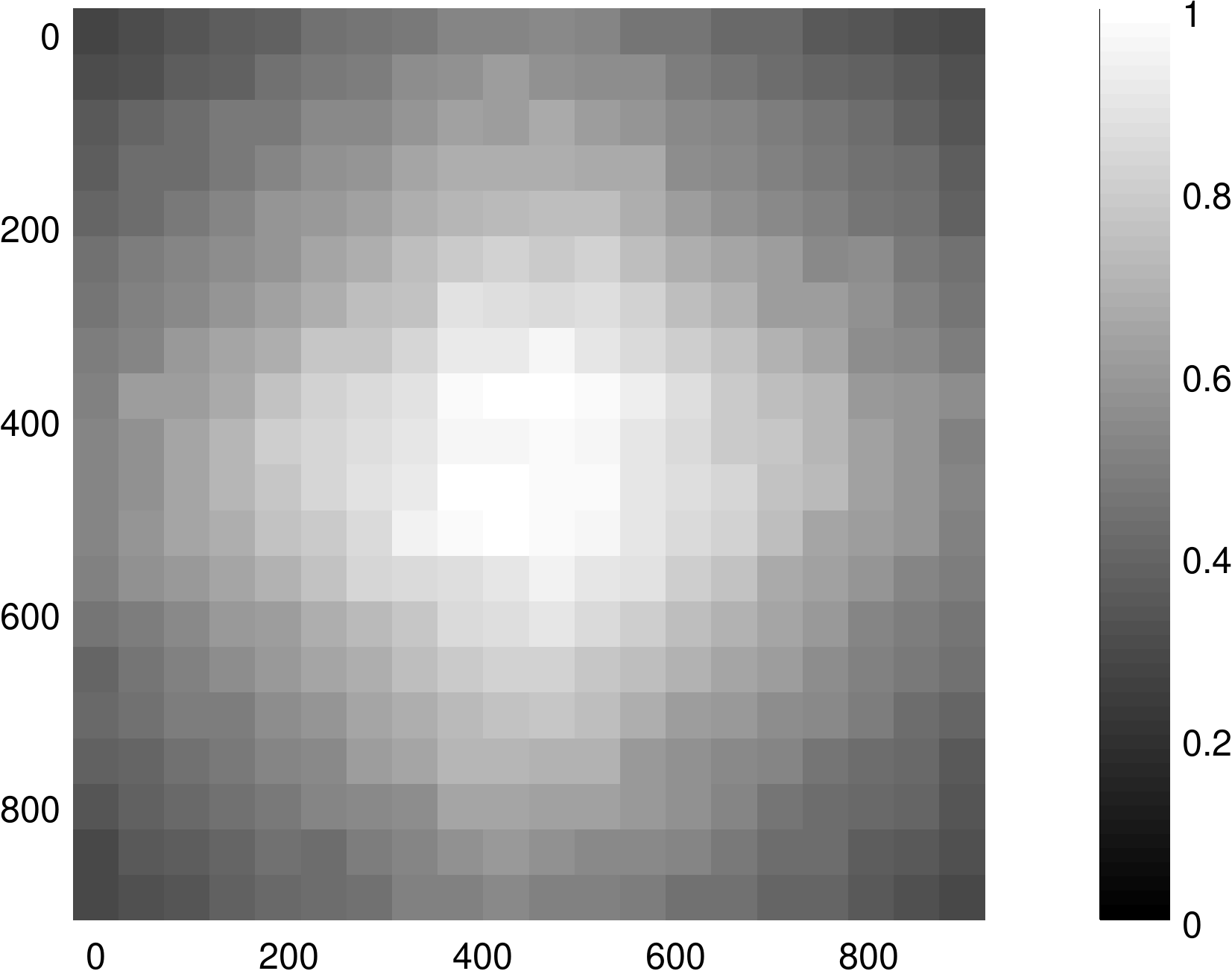}
 \caption{ Probability as a function of \x-\y~position (mm) that an alpha particle from a $^{218}$Po decay at
 the central cathode is fully contained within the detector volume.  Results obtained from simulation.}

 \label{fig:PoXY}
\end{figure}

The derivation above measures the probability of detecting a particular $^{222}$Rn \itshape given that the subsequent $^{218}$Po alpha is also observed.  \upshape 
However, because the focus lies on the two consecutive decays from the same parent $^{222}$Rn atom, there is necessarily a 
spatial correlation between the two events.  
This is particularly true here because the $^{218}$Po atom is positively charged about $80\%$ of the time (see Section \ref{sec:neutral}) 
and follows the electric field lines to the central cathode with negligible diffusion, thus maintaining the $x$-$y$ position of the atom.  
This spatial correlation is important to consider because alpha 
tracks which originate near the center of the detector are much more likely to remain contained than tracks which originate near a fiducial boundary.  
We account for this correlation by reusing the simulation which we used in Section \ref{sec:ContFrac} to calculate
that 0.121 of the $^{222}$Rn decays produce GPCCs.  
Here we simulate $^{218}$Po alpha decays from the central cathode and record the original \x-\y~position of 
the decay when the track is contained - the results of this are shown in Figure \ref{fig:PoXY}.
Finally, we once again model alphas from $^{222}$Rn decays but use the distribution from Figure \ref{fig:PoXY}, rather than a homogeneous distribution, to 
choose the starting position of the atom.  
This results in a fraction of 0.130, rather than 0.121, for the production of GPCCs from $^{222}$Rn decays for which the subsequent $^{218}$Po alpha is later observed.  
This indicates that the spatial correlation shifts this efficiency measurement high by $7.4\%$.

After correcting for this shift, the efficiency calculated by this technique for $^{222}$Rn falls to 0.054.  
This result is $4\%$ higher than the efficiency of 0.052 derived using the method described in Sections \ref{sec:ContFrac}-\ref{sec:angles}.  
In comparing the efficiencies obtained by these two different methods for both contained 
and GPCC populations of $^{222}$Rn alpha tracks, the largest disagreement is $10\%$.  
This is conservatively taken to be the systematic uncertainty in the absolute efficiency for observing and identifying alpha particle tracks.  

The time correlation technique can only be used in DRIFT-IId to measure the efficiency of detecting alpha particles from $^{222}$Rn decays;
however it may be useful to other low-background experiments that have backgrounds caused by radioactive decay chains with short half-lives.  
The uranium isotopes have half-lives that are too long, and the polonium atoms are complicated by being present both in the gas and on the central cathode.  
The efficiencies used in this document, presented in Table \ref{tab:Efficiencies}, 
 are calculated using the method described in Sections \ref{sec:ContFrac}-\ref{sec:angles}.
The assumptions from this method, however, are validated by this second, entirely independent efficiency calculation by time correlation.

\section{Thin-Film Cathode}
\label{sec:ThinFilm}

With the alpha analysis presented above we now have the tools needed to characterize the contaminations and backgrounds due to alpha decays at the central cathode. 
Here we also introduce the thin film cathode as a method of reducing the backgrounds produced by these decays.  

As described in Section \ref{sec:RPRs} and Figure \ref{fig:Backgrounds}, an RPR occurs when an alpha decay at the cathode is misidentified 
as a nuclear recoil because the alpha particle is lost in the cathode material.  In the case when the alpha also enters the gas volume, 
it is used to tag and veto the event; such events are classified as ``tagged-RPRs.'' 

The original 20~\micron~wire cathode had a large RPR background, resulting from a significant fraction of un-tagged RPRs. 
The most  common isotopes in the $^{222}$Rn chain to decay at the cathode are $^{218}$Po ($^{214}$Po), 
 which decay to produce 6.00 (7.69) MeV alpha particles. 
These have a path length of 12.4 (17.8) \micron~in stainless steel \cite{SRIM}. 
By  geometrical calculation, this corresponds to 34\% (24\%) of the alpha particles lost in the wire, producing RPR background events. 
A thinner central cathode that is almost transparent to alpha particles would greatly increase the probability that an 
 RPR is tagged by the alpha track, essentially eliminating the RPR.  

The most promising material candidate is $0.9$~\si{\micro \meter} biaxially-oriented polyethylene terephthalate (mylar) with a $300-500$~\AA~layer of aluminum 
vacuum-deposited on each side for electrical conductivity.  This material reduces the likelihood for an alpha particle to be fully absorbed from around
$30\%$ to $1\%$ or less. 
The DRIFT collaboration installed a new central cathode constructed of this thin-film in March 2010 and collected 26.22 live-time days of shielded data.  

\begin{figure}[t]
  \centering
  \begin{minipage}[t]{0.475\textwidth}
    \centering
    \includegraphics[width=\textwidth]{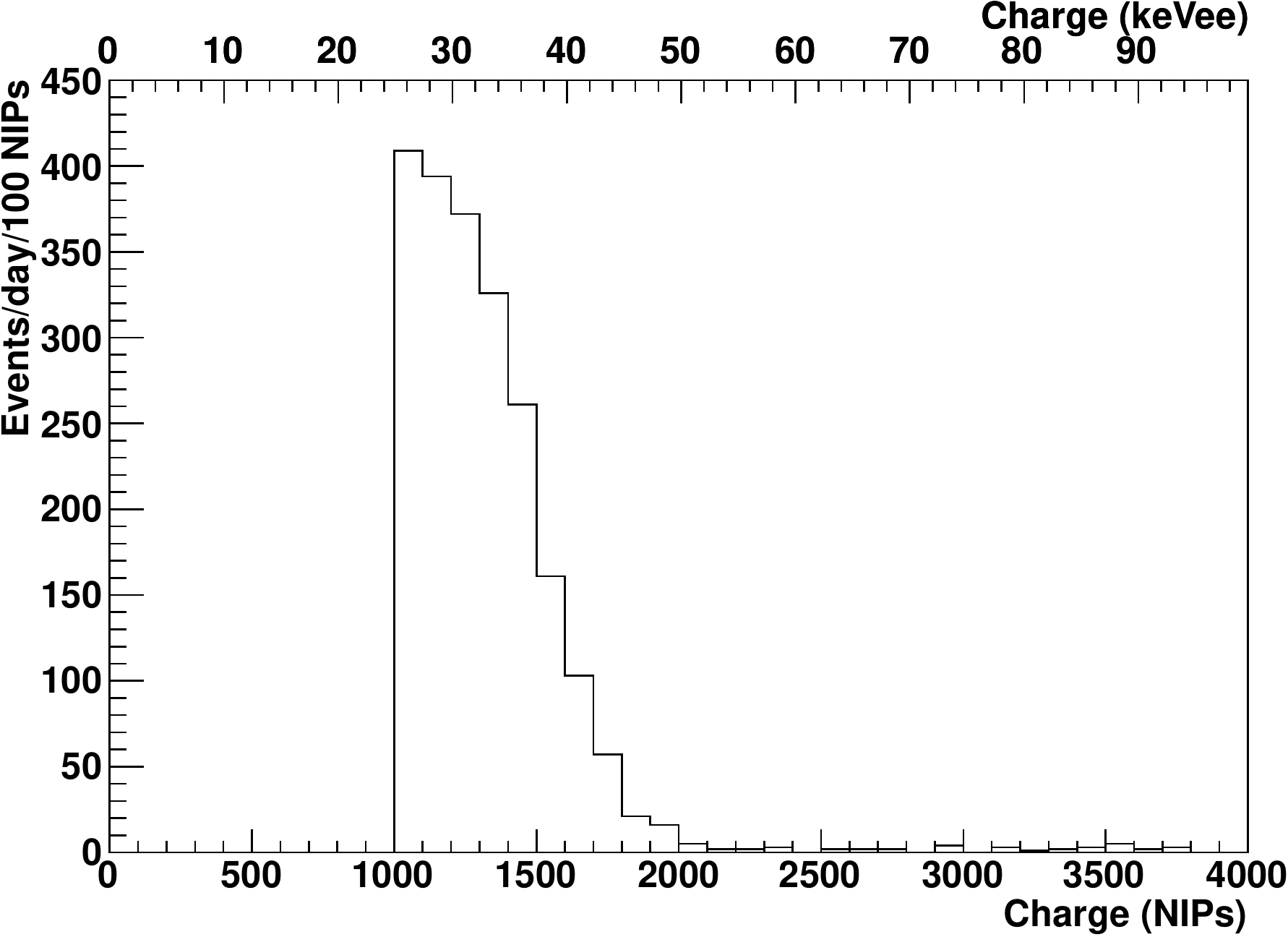}
    \captionof{figure}{ Measured energy distribution of background events from the wire cathode. }
    \label{fig:WireRPR}
  \end{minipage}%
  \hfill
  \begin{minipage}[t]{0.475\textwidth}
    \centering
    \includegraphics[width=\textwidth]{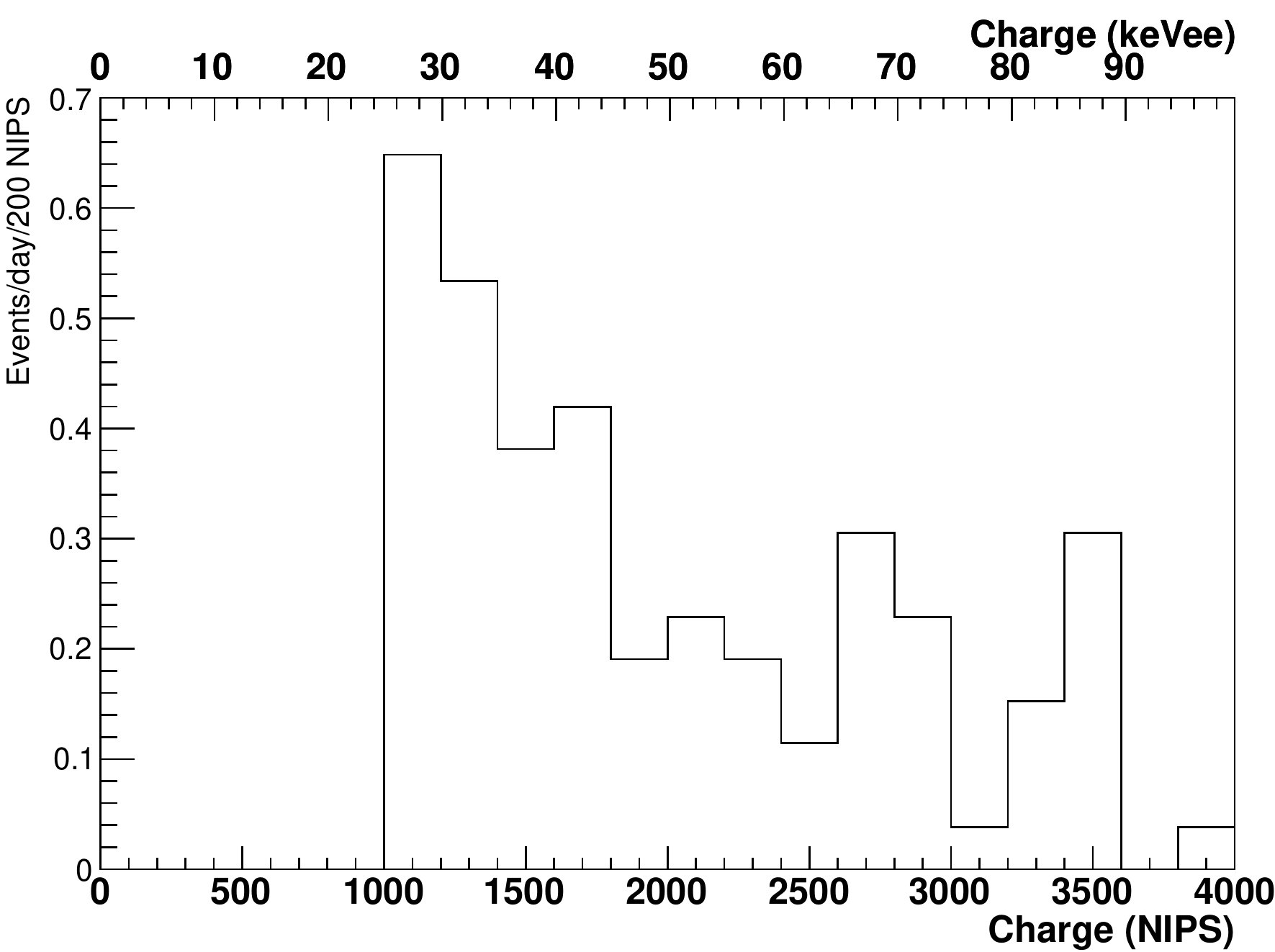}
    \captionof{figure}{ Measured energy distribution of background events from the thin-film cathode. }
    \label{fig:FilmRPR}
  \end{minipage}
\end{figure}

The installation of the thin-film cathode led to an immediate reduction in background rate, however, it was only a factor 19 instead of the factor 30 predicted 
above from geometry.  In addition, the distribution of background events also changed in unexpected ways.  The population of RPRs was easily recognizable, 
if reduced, but the background events from the central cathode now extended past the upper limit in energy for RPRs (see Figs \ref{fig:WireRPR} and 
\ref{fig:FilmRPR}).  These were suspected to be Low-Energy Alphas (LEAs).  

\subsection{Low-Energy Alphas}
\label{sec:LEA}

As shown in Figure \ref{fig:Backgrounds}, an LEA is an event where only a short segment of an alpha track projects out of the cathode into 
the gas on one side of the detector volume, with no ionization appearing on the other side. Such an event may be misidentified as a nuclear
recoil only if the alpha range in the gas is $\lesssim 10$~mm.  
This can be thought of as the inverse of an RPR event. 
There, the alpha must be absorbed to prevent tagging of RPRs, whereas for an LEA the recoil
must be entirely absorbed in the cathode material in order for charge to fall on only one side of the detector.\footnote{DRIFT's
WIMP analysis checks for ionization on the opposite side of the detector and coincident in time to exclude exactly this type of event.}

\begin{figure}[t]
 \centering
 \includegraphics[width=0.8\textwidth]{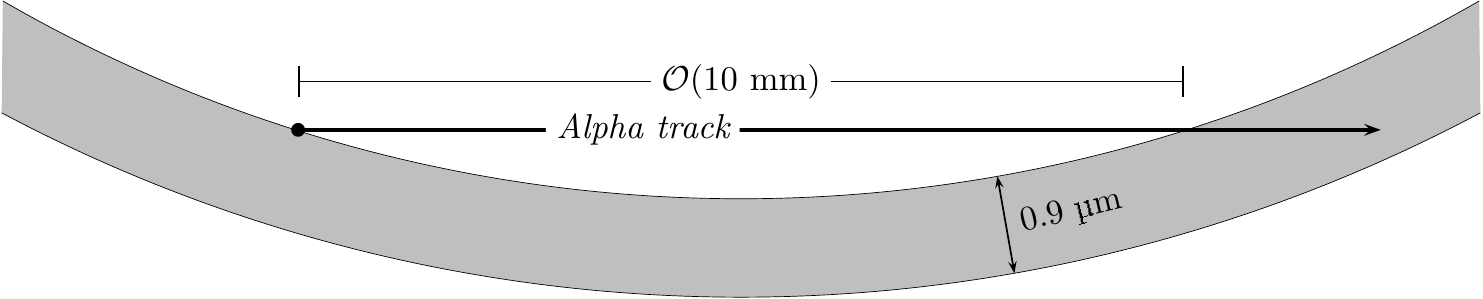}
 \caption{ Diagram of how a low-energy alpha (LEA) may be produced due to ripples across the thin-film cathode.}
 \label{fig:LEA}
\end{figure}

The alpha particle, which initially carries far too much energy to be mistaken for a nuclear recoil, must lose most of its energy by
passing through the cathode material along its path.  
While there are many potential geometries for this to occur, we present two here to illustrate possibilities.  
Figure \ref{fig:LEA} shows the emission of an alpha particle at a shallow angle from the surface of the thin-film cathode.  
The alpha particle is oriented into the gas while the daughter atom embeds itself in the cathode, producing no ionization in the gas.  
Because the thin-film cathode is not perfectly flat, the alpha particle may cross a trough and later re-enter
the cathode, thus producing only a short track of ionization in the gas.  Similarly, Figure \ref{fig:Backgrounds} shows an LEA created
by the emission of an alpha particle from an atom buried below the surface of the cathode.  The daughter atom never exits the film while the
alpha particle, traveling nearly parallel to the plane of the cathode, exits near the end of its track to deposit only a small amount of 
ionization before ranging out in the gas.

The buried atoms necessary for the second process were common in this DRIFT-IId cathode.  Uranium, present in the aluminum layer of the 
thin-film cathode (see Section \ref{sec:UCont}), is typically buried.  
Radon progeny can also end up being buried in the cathode material.  While $^{218}$Po, produced by $^{222}$Rn decays in the gas, 
can only be deposited onto the cathode surface, its
daughters by alpha decay can be embedded.  After $^{218}$Po decays at the surface, the resultant $^{214}$Pb atom will be directed into the film half of the time. 
With 112~keV of kinetic energy, it can embed itself up to 400\AA~deep into the aluminum layer of the film.  This atom, which rapidly beta-decays
into the alpha-producing $^{214}$Po, is now buried in the cathode and can produce an LEA.  

While RPRs have a distinct energy spectrum rarely creating more than 2000 
NIPs,\footnote{2000 NIPs is $3\sigma$ above the mean energy for $^{210}$Pb recoils, which are the highest-energy RPRs in DRIFT.}
(see Figure \ref{fig:WireRPR}) LEAs do not have such a limitation.  
They instead exhibit a nearly flat energy distribution across DRIFT's entire signal region.  
This feature allows populations of LEAs to be  distinguished from RPRs.  
LEAs are present in data taken with both the wire and the thin-film cathodes; however with the wire cathode the RPR rate swamps the LEAs \cite{Burgos2007}.  
The thin-film considerably reduces rate of RPRs such that the LEAs become a significant contributor to the background rate of the experiment.

With the discovery of the LEAs in the thin-film cathode data, the analysis focused on determining the decaying isotopes and their locations. 
Figure \ref{fig:Cont} begins to provide a method for determining where a type of alpha-decay occurred. 
These data were taken with the thin film cathode in place and show large uranium contributions. 
It was suspected that these were from the cathode; however, using alpha range spectroscopy it is possible to pin down the location of the radio impurities.

\section{Contamination Location}
\label{sec:Location}

In Section \ref{sec:RPRs}, Equation \ref{eq:FoM}, we defined a figure of merit to assess the efficacy of different cathodes for rejecting alpha-decay 
backgrounds at the cathode. 
The denominator of this quantity is the sum of all alpha decays which occur on or in the central cathode.  
While some populations of alpha-decaying contaminants, such as uranium isotopes, are present only at the cathode, 
others, for example $^{222}$Rn, are present in the gas and do not contribute backgrounds in DRIFT-IId.  
The contaminants of interest, as identified in Figures \ref{fig:GPCC} and \ref{fig:Cont}, are radon's daughters $^{214}$Po and $^{218}$Po
(Section \ref{sec:neutral}) as well as two isotopes of uranium (Section \ref{sec:UCont}). 
The Po atoms present on the central cathode contribute to the figure of merit, while some fraction is neutrally-charged and remains in the gas.   
Below we show the measurement of neutral fractions for $^{218}$Po and $^{214}$Po fractions, from which we deduce their contribution to the alpha-decays at the cathode.

\subsection{Polonium Neutral Fraction}
\label{sec:neutral}

When an atom undergoes an alpha decay, the recoiling nucleus loses its outer shell of electrons and then, as it comes to rest, 
regains some or all of them \cite{Hopke1996}.  
If it is positively charged when it comes to a stop, it will follow the electric field lines and electrodeposit onto the central cathode.  
If instead it is neutral, it will remain in the gas, potentially producing a GPCC if it does later alpha-decay.  

\begin{table}
 \centering
  \begin{tabular}{l|c|c}
  Isotope & Absolute Count & Neutral Fraction ($n$)\\
  \hline
  $^{222}$Rn & $96000\pm10000$ & $1.00$ \\
  $^{218}$Po & $22200\pm2500$ & $0.23\pm0.3$\\
  $^{214}$Po & $0^{+200}_{-0}$ & $0.00^{+0.002}_{-0}$\\
  \end{tabular}
 \caption{Neutral fraction of various isotopes.  }
 \label{tab:neutral}
\end{table}

Alpha particles emitted by a particular isotope are selected from the population of GPCC alpha tracks by their lengths, and counted.  
The absolute count of decays by this isotope in the gas is calculated by dividing this observed number by the identification efficiency for that isotope, 
described in Section \ref{sec:efficiency}.  
Due to the short half-lives of the Po and Bi isotopes, the $^{214}$Po and $^{218}$Po are in equilibrium with 
$^{222}$Rn.\footnote{Gas flow reduces Po concentration by less than $0.1\%$}
The neutral fraction of each polonium isotope is therefore the ratio of Po decays in the gas to $^{222}$Rn decays in the gas.  
The absolute counts and resultant neutral fractions are provided in Table \ref{tab:neutral}.  
Decay counts of polonium atoms on the cathode are obtained by using this neutral fraction and extrapolating from the known absolute count of $^{222}$Rn decays.  
This is used to calculate the figure of merit below in Section \ref{sec:QuantCath}.  

There is a considerable difference in the measured neutral fraction between $^{218}$Po ($23\%$) and $^{214}$Po ($0\%$).  This is likely to be due to the
chain of radioactive decays between the $^{218}$Po and $^{214}$Po alpha emissions.  $^{218}$Po is created by alpha emission of $^{222}$Rn in the gas.
When $^{218}$Po decays, it produces an alpha particle and a $^{214}$Pb 
atom which is either neutral or positively charged.  
An atom which is neutral will remain in the gas until, with $t_{1/2}=26.8 $ min, it undergoes a $\beta$ decay to produce $^{214}$Bi.  
By emitting an electron, the resultant bismuth atom naturally obtains a positive charge, 
and it is carried to the cathode by the electric field.  

\subsection{Uranium}
\label{sec:UCont}

There are generally three possibilities for the location of the contamination in DRIFT-IId: the gas, the cathode, or the MWPCs.  
There is no mechanism by which uranium should enter the gas,\footnote{This is supported by the absence of uranium GPCCs in DRIFT-IId.}
 which leaves the MWPCs and the central cathode. These can be distinguished by using the vector direction of the alpha particle track.  

\begin{figure}[t]
 \centering
 \includegraphics[width=0.6\textwidth]{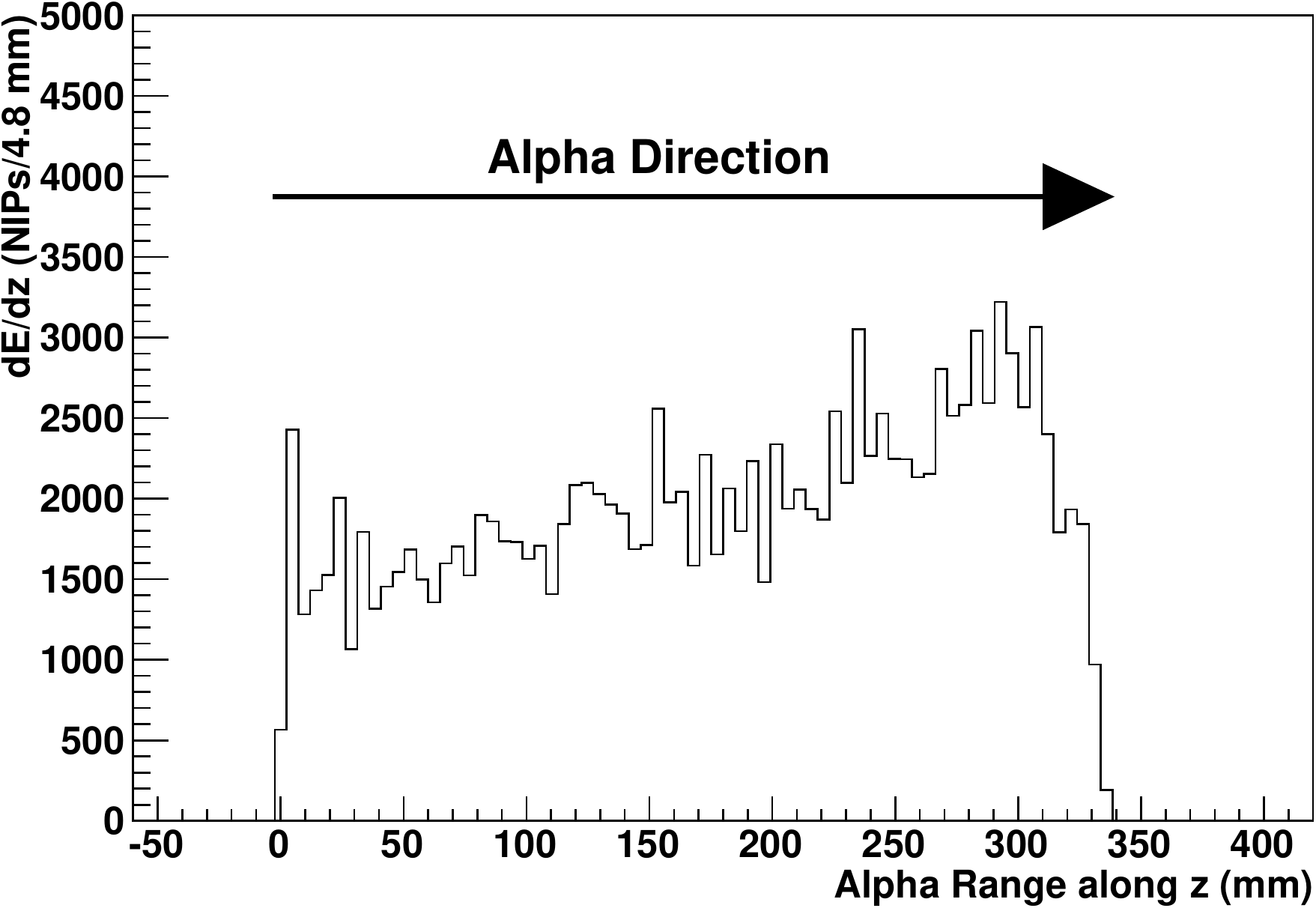}
 \caption{ The ionization produced by a 5.49 MeV alpha particle projected onto the \z~ axis.  
 The direction of the alpha particle is measured by this asymmetry.  }
 \label{fig:Bragg}
\end{figure}

\begin{figure}[t]
  \centering
  \begin{minipage}[t]{0.475\textwidth}
    \centering
    \includegraphics[width=\textwidth]{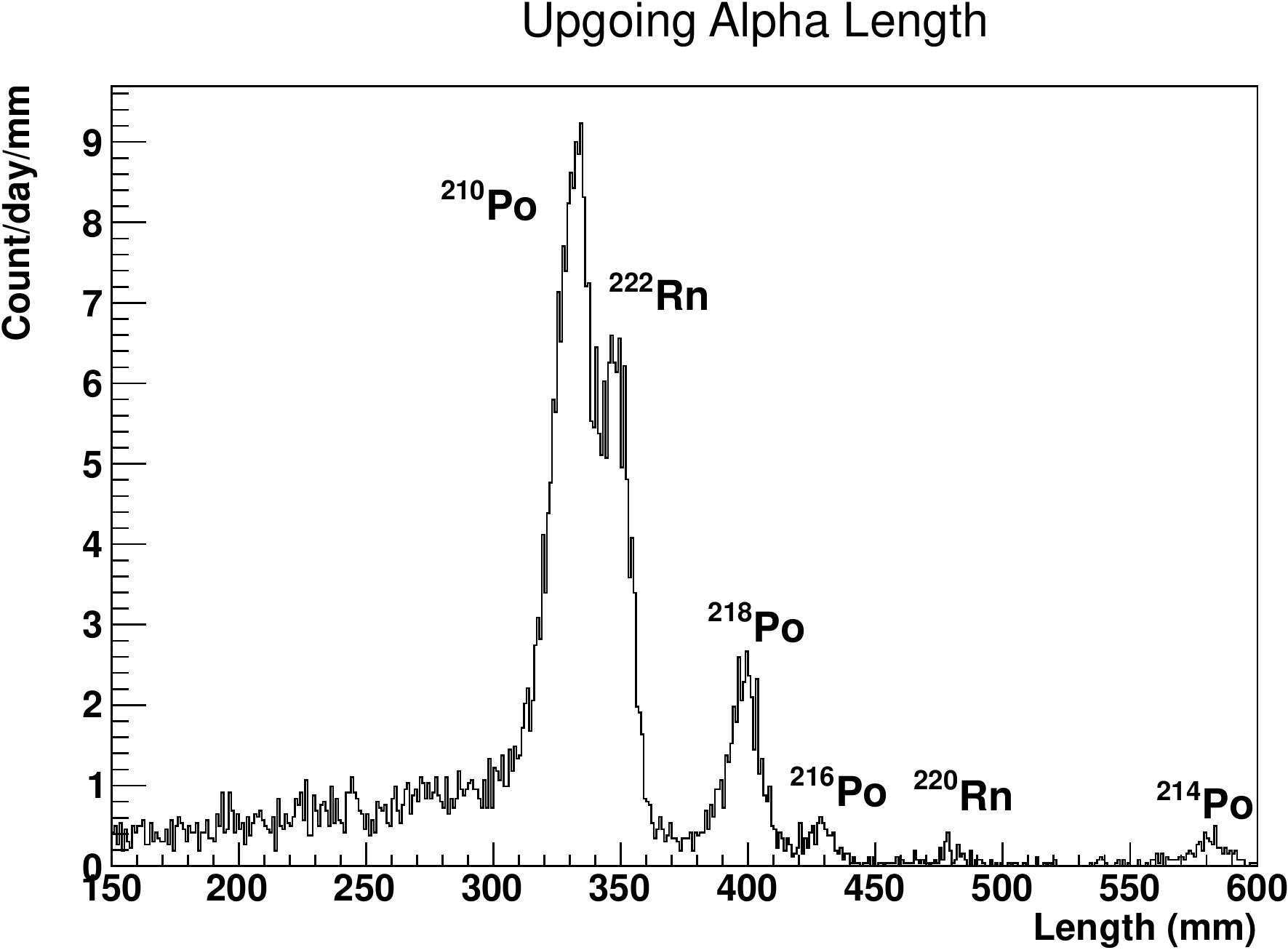}
    \captionof{figure}{ Up-going alpha tracks. }
    \label{fig:FilmUp}
  \end{minipage}%
  \begin{minipage}[t]{0.05\textwidth}
  \hspace{\textwidth}
  \end{minipage}%
  \begin{minipage}[t]{0.475\textwidth}
    \centering
    \includegraphics[width=\textwidth]{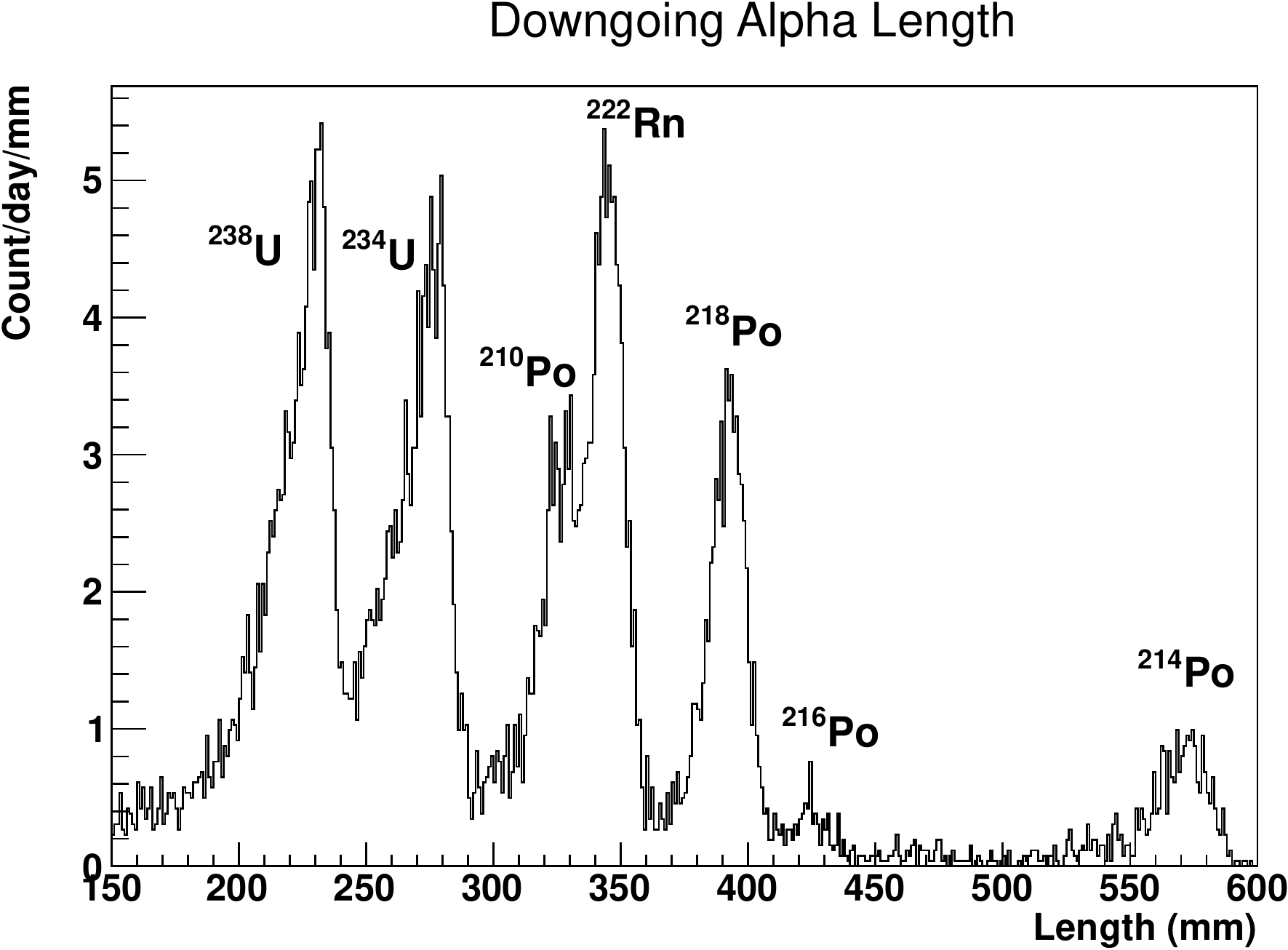}
    \captionof{figure}{ Down-going alpha range spectrum.\protect\footnotemark }
    \label{fig:FilmDown}
  \end{minipage}
\end{figure}

Alpha particle tracks exhibit a Bragg peak in which they produce more ionization near the end of the track (see Figure \ref{fig:Bragg}).  
This is used to distinguish between the origin and the end of a track, and to then categorize it as either traveling toward the central cathode (``up-going'')
or away from the central cathode (``down-going'').  Figures \ref{fig:FilmUp} and \ref{fig:FilmDown} show the length distribution of up-going
and down-going contained alpha tracks, respectively.  These reveal that all of the
uranium is on or in the central cathode because every alpha emitted by a uranium decay is oriented away from the central cathode.  Alpha particles
which originate from decays in the gas, like those due to the decay of $^{222}$Rn, are present in equal quantities in both up-going and down-going distributions.  
Finally, isotopes that are present on both the MWPCs and the central cathode are also present in both populations, although typically in unequal
quantities.  For example, alpha particles from the decay of $^{210}$Po are almost entirely up-going,  indicating contamination of the MWPCs.\footnote{
A dirty MWPC can be identified by further dividing the up-going population into decays which occur on the left and right sides of the detector.}

\begin{figure}[h] 
  \begin{subfigure}[t]{0.47\linewidth}
    \centering
    \includegraphics[width=0.95\linewidth]{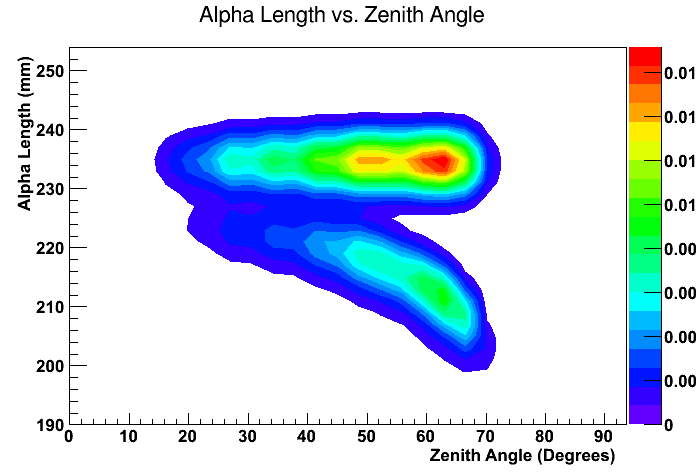} 
    \caption{Uranium in aluminum model.} 
    \label{fig:ModelAl} 
    \vspace{4ex}
  \end{subfigure}
	\hfill
  \begin{subfigure}[t]{0.47\linewidth}
    \centering
    \includegraphics[width=0.95\linewidth]{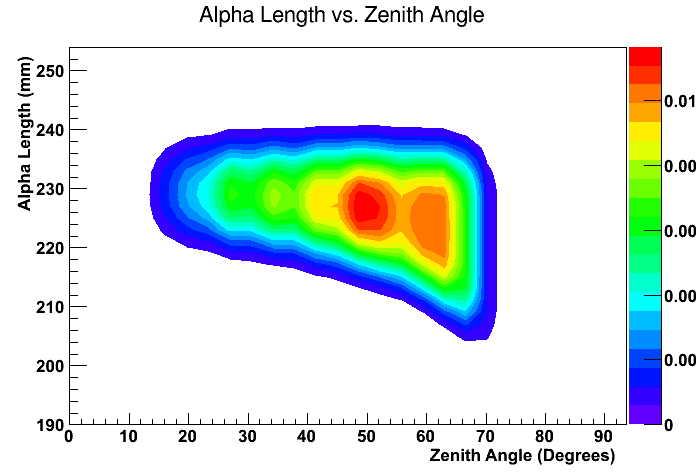} 
    \caption{Uranium in Mylar model.} 
    \label{fig:ModelFilm} 
    \vspace{4ex}
  \end{subfigure} 
  \begin{subfigure}[t]{0.47\linewidth}
    \centering
    \includegraphics[width=0.95\linewidth]{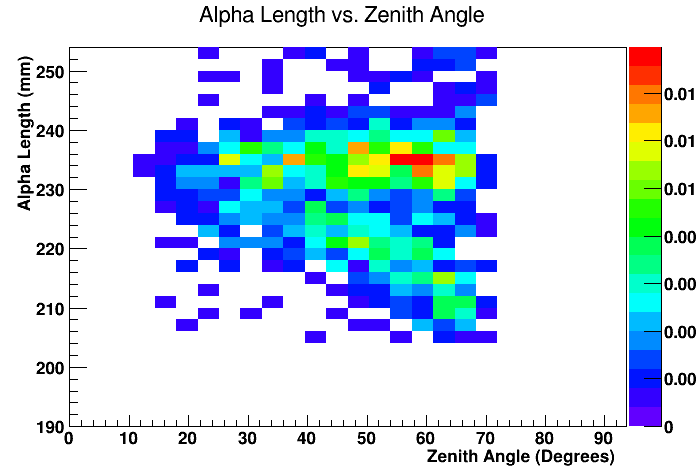} 
    \caption{Data from DRIFT.} 
    \label{fig:Data} 
  \end{subfigure}
	\hfill
  \begin{subfigure}[t]{0.47\linewidth}
    \centering
    \includegraphics[width=0.95\linewidth]{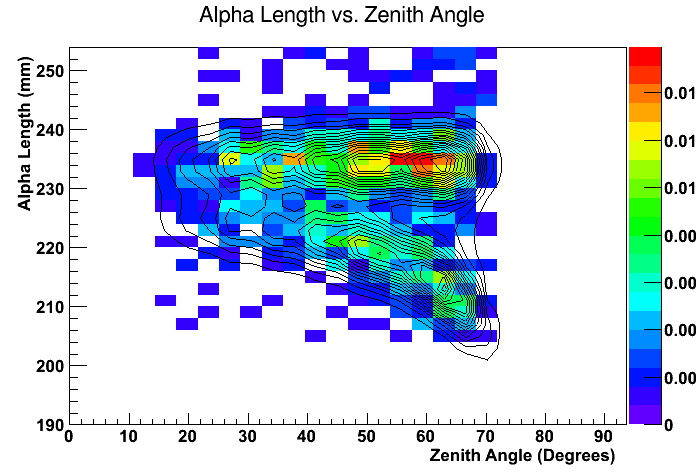} 
    \caption{The data (histogram) overlaid with the best fit model (contour).} 
    \label{fig:ModelOverlay} 
  \end{subfigure} 
  \caption{The shapes of the range vs. angle distribution for uranium indicate that the contamination is in the aluminum rather than the mylar.  }
  \label{fig:Models} 
\end{figure}

\footnotetext{The asymmetry of the uranium distributions is due to the alpha particles which lose some energy as they 
    pass through some of the thin-film cathode and therefore produce shorter \tracks~in the gas (consider Figures \ref{fig:ModelAl}-\ref{fig:ModelOverlay}).}

\begin{figure}[h]
 \centering
 \includegraphics[width=0.65\textwidth]{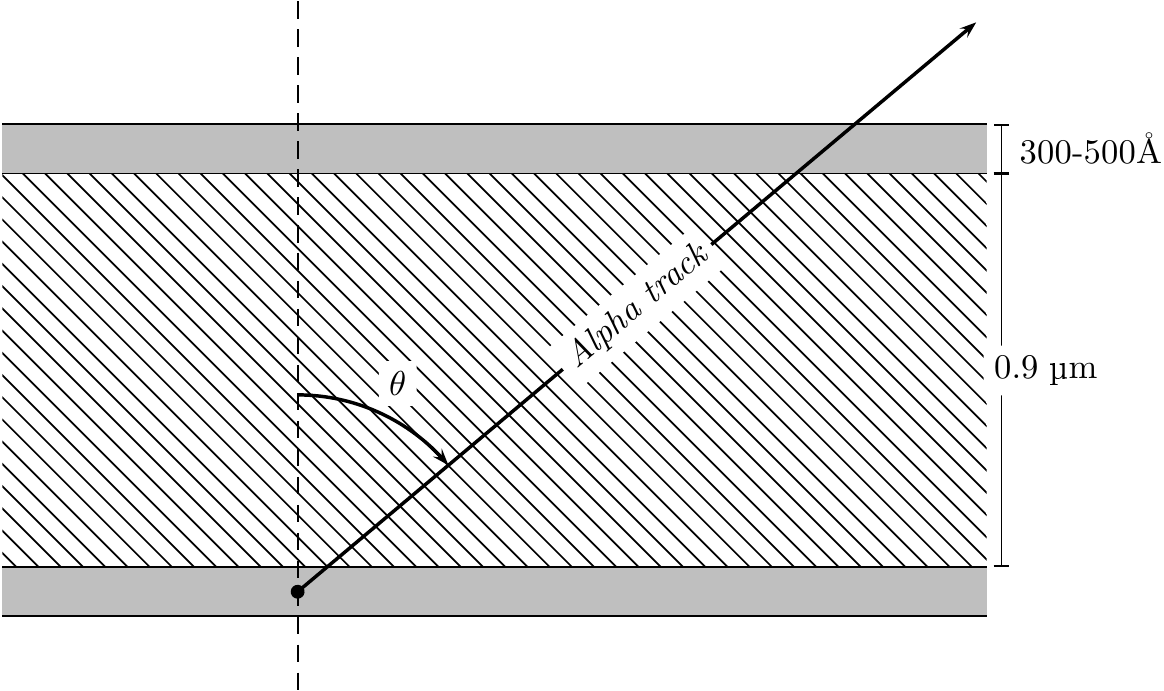}
 \caption{ An Alpha from a Uranium decay in the aluminum layer, passing through the mylar.  As the angle relative to the zenith ($\theta$) increases, 
 the distance traveled through the mylar increases, shortening the length of the Alpha in the gas.  }
 \label{fig:Schematic}
\end{figure}

This uranium contamination of the central cathode can be quantified.  
Using the efficiency calculations of Section \ref{sec:efficiency} we obtain an absolute count of uranium decays from the central cathode and, from this,
an absolute number of atoms of each isotope.  Present in the 1.1~g of thin film within the cathode's fiducial area, this corresponds to a contamination of 
$61.8\pm0.6$~ppt $^{234}$U and $777\pm15$~ppb $^{238}$U.  
This value is consistent with an independent assay done at SNOLAB \cite{SNOLAB}, which measured $2140\pm1550$~ppb $^{238}$U.  
With contamination on the thin-film cathode, DRIFT-IId has the advantage of measuring the uranium atom's decay directly, rather than the decay of a daughter isotope,
and with high ($\approx 35\%$) efficiency.

The uranium contamination, now known to be located only at the central cathode, can be removed by replacing this component with a cleaner version.  
To do this, however, it is important to know if the uranium is present in the aluminum layer, the mylar layer, or both.  
The location of the uranium can be pinpointed by examining the relationship between the alpha angle ($\theta$) and the alpha length in the gas.  
For alphas originating at or near to the surface of the cathode and exiting directly into the gas without passing through the cathode, the length is 
nearly independent of $\theta$. 
If, instead, the surface-originating alpha passes through the higher-density cathode before entering into the low-density the gas on the opposite side (Figure \ref{fig:Schematic}), 
the length will decrease by $\cos(\theta)^{-1}$. 
Thus, if uranium contamination is restricted to the thin surface layers of aluminum, a bimodal length vs. $\theta$ distribution would be expected. 
If instead the contamination is restricted to the mylar, this distribution will be unimodal.

To quantify this, the length vs. $\theta$ distribution was modeled with the uranium contamination being either in the aluminum or in the mylar. 
Alphas from uranium decays were generated isotropically with absorption coefficients obtained from SRIM \cite{SRIM}, and their final lengths were 
convolved with a Gaussian to account for longitudinal straggling and measurement error. 
The resulting length vs. $\theta$ distribution is also modulated by DRIFT's efficiency over $\theta$ for detection (Section \ref{sec:angles}). 
For the model corresponding to uranium in the aluminum layer, another angle-dependent efficiency is included to account for 
the probability that the recoiling thorium atom leaves the film and ``vetoes'' the alpha track. 
The resultant models are shown in Figure \ref{fig:ModelAl} and \ref{fig:ModelFilm}.

The data shown in Figure \ref{fig:Data} clearly shows the bimodal distribution expected from the aluminum-only contamination model. 
A 2-parameter fit is performed matching a superposition of the two model distributions to that obtained from the data. 
In the best fit model (Figure \ref{fig:ModelOverlay}), 85$\%$ of the uranium contamination is in the aluminum and $15\%$ is in the mylar. 
With this, we were able to unambiguously pinpoint the location of the majority of the uranium contamination to the $\approx400$~\AA~layers of aluminum, 
allowing us to focus on that component for the next version of the thin-film cathode.

\section{Clean Thin-Film Cathode}
\label{sec:Clean}

\begin{figure}[ht]
  \centering
  \begin{minipage}[t]{0.475\textwidth}
    \centering
 \includegraphics[width=0.9\textwidth]{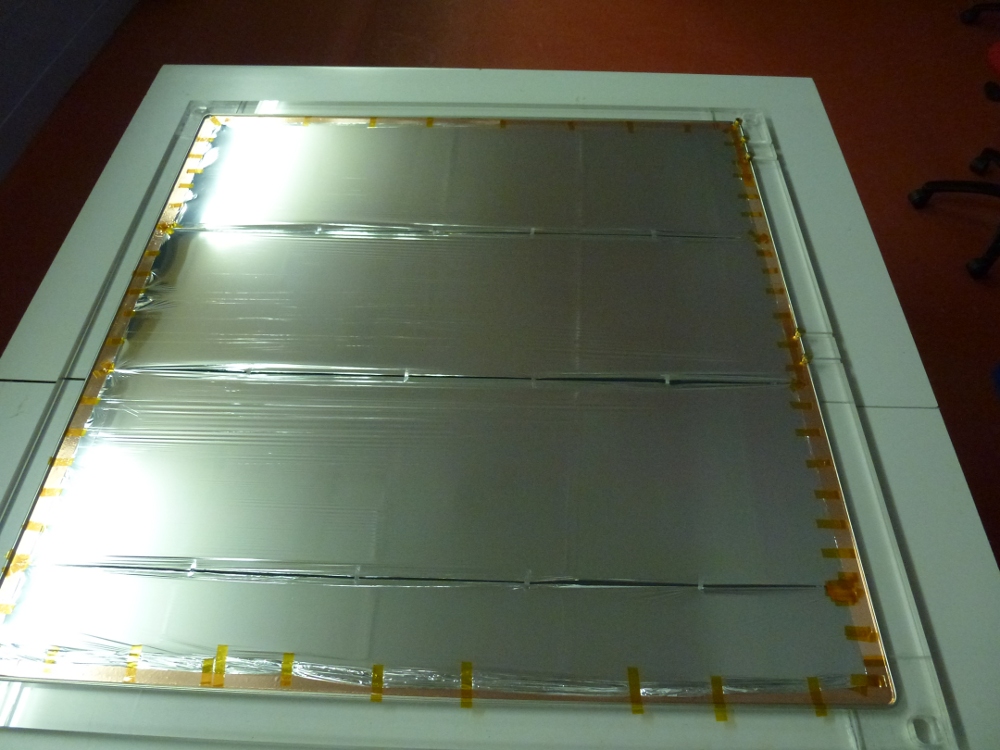}
 \captionof{figure}{ Photograph of the DRIFT-IId central cathode with a low-uranium content thin-film cathode freshly installed.  }
    \label{fig:CathodeImage}
  \end{minipage}%
  \begin{minipage}[t]{0.05\textwidth}
  \hspace{\textwidth}
  \end{minipage}%
  \begin{minipage}[t]{0.475\textwidth}
    \centering
    \includegraphics[width=\textwidth]{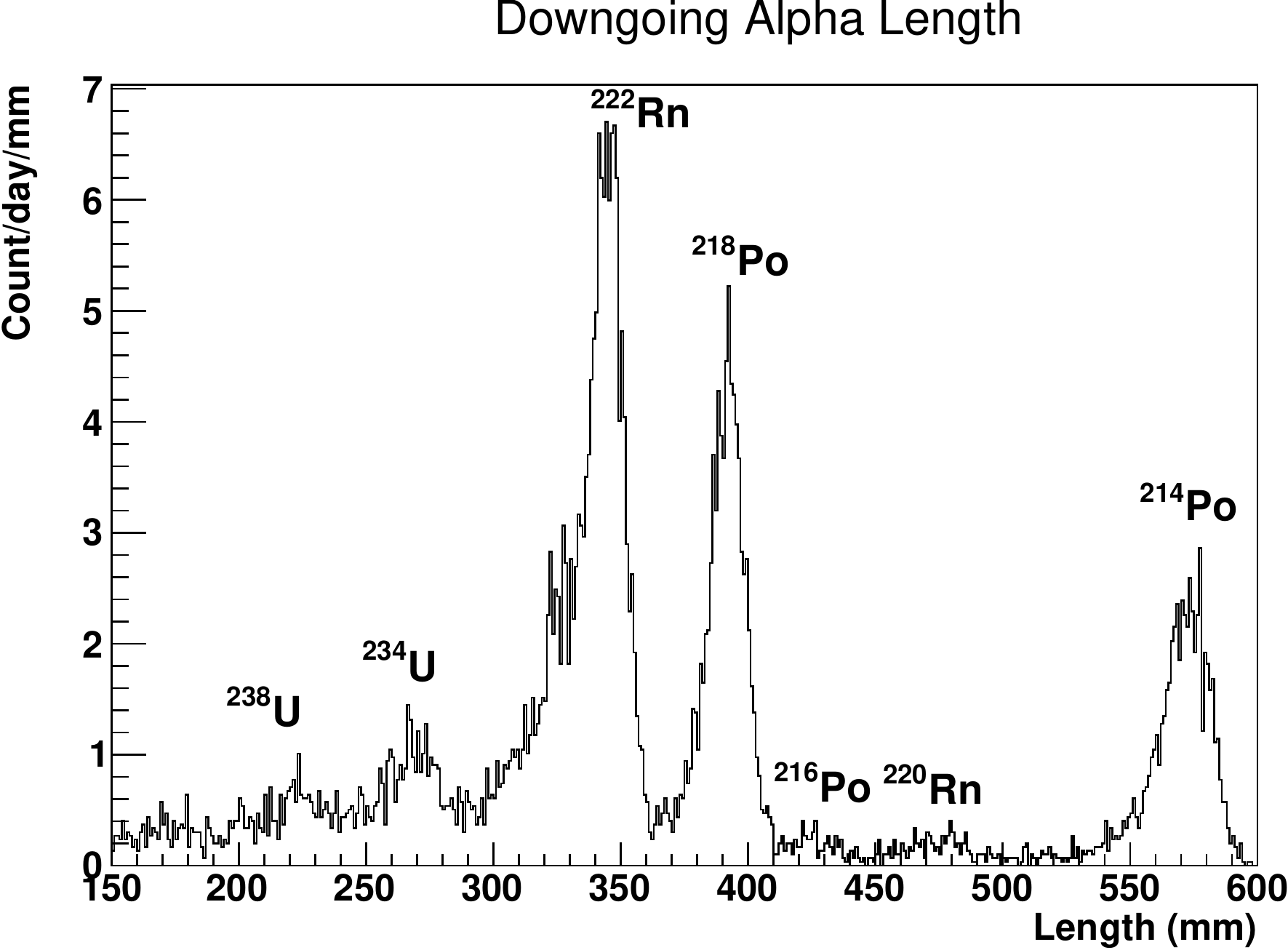}
    \captionof{figure}{ Downgoing alpha tracks from the clean-film runs. }
    \label{fig:CleanFilmDown}
  \end{minipage}
\end{figure}

Following the confirmation that the uranium contamination was primarily in the aluminum layer, the DRIFT collaboration manufactured a new set of aluminized 
mylar cathode panels, this time using ultra-pure aluminum.  This new cathode was installed in April of 2012 and DRIFT collected 29.69 livetime-days of
shielded data.  The $^{238}$U contamination level dropped from $777\pm15$ to $73\pm2$ ppb (see Table \ref{tab:UCont}), confirming the measurement
in Section \ref{sec:UCont} which located the majority of the contamination in the aluminum layer.  The background rate also dropped further - these results
are presented and discussed in detail in Section \ref{sec:QuantCath}.

\begin{table}[t]
 \centering
  \begin{tabular}{l|c|c}
  Isotope & Dirty & Clean \\
  \hline
  $^{234}$U & $61.8\pm0.6$ ppt & $ 3.3\pm0.1$ ppt \\
  $^{238}$U & $777\pm15$ ppb & $73\pm2$ ppb\\
  \end{tabular}
 \caption{Uranium contamination levels during different runs. Uncertainties shown are statistical; all values have an associated 10\% systematic
 uncertainty from the alpha detection efficiency.  }
 \label{tab:UCont}
\end{table}

Figure \ref{fig:CathodeImage} shows the thin-film installed on a cathode frame.  The aluminized mylar has only been available in 1-ft widths, so 4 strips
(three full and one narrow) have been used to span the full 1~m width of the cathode frame.  Each panel is attached only on its short end, so it
is taut longitudinally but not laterally.  This results in ripples present in the film, with peaks running lengthwise along each panel.  
In this version an attempt was made to reduce the impact of these ripples in the film on LEA production by flattening the panels better.  In between
each pair of panels four small ($\approx 1$ cm$^2$) tabs were used to connect them to allow the application of lateral tension as well.  This
reduced the ripples, but did not eliminate them.  
The area of these tabs is $\approx0.05\%$ of the total cathode area which, due to the increased thickness in these regions, is expected to increase the overall
background rate by around 3\%.

\section{Texturized Thin-Film}
\begin{figure}[t]
  \begin{minipage}[ht]{0.475\textwidth}
    \centering
    \includegraphics[width=\textwidth]{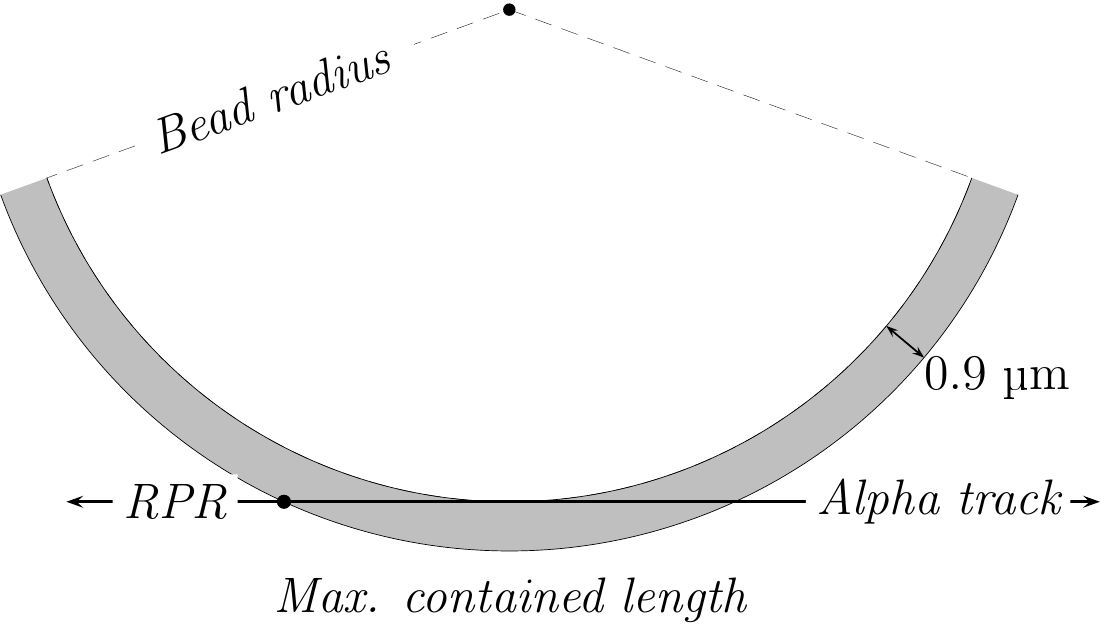}
    \caption{ Schematic showing the motivation behind texturizing the thin-film cathode.  Here the cathode is formed into a series of 
    hemispheres such that the longest straight-line path (shown) is shorter than the path length of an alpha particle track.
    An alpha particle cannot be fully absorbed by the cathode so it always deposits ionization in the gas, providing a tag to reject RPR events.   }
    \label{fig:TexSimple}
  \end{minipage}%
  \hfill
  \begin{minipage}[ht]{0.475\textwidth}
    \centering
    \includegraphics[width=0.9\textwidth]{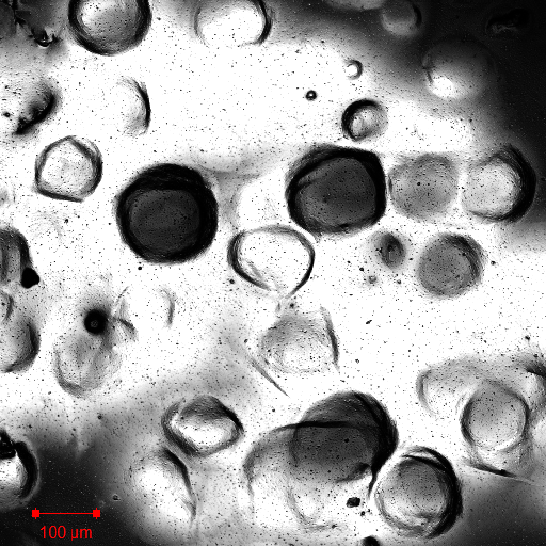}
    \caption{ A confocal microscope image of the texturized thin-film.   }
    \label{fig:FilmImage}
  \end{minipage}
\end{figure}

Any reduction of the probability for an alpha to be fully absorbed in the cathode will result directly in a reduction of background rate in DRIFT-IId.  
While handling even thinner cathodes is impractical, it is possible to imprint the thin-film with a 3-dimensional pattern to guarantee that 
there are no straight-line paths in the film long enough to absorb an entire alpha track (see Figure \ref{fig:TexSimple}).  There are industrial
applications which micro-texturize thin films, but not films thinner than 25 \micron.  

Using techniques to be described fully in a later document, it is possible to texturize 0.9 \micron~aluminized mylar by impacting it with uniformly sized glass
beads (see Figure \ref{fig:FilmImage}).  Due to the curvature of an impact pit from a spherical bead, a diameter $\leq300$ \micron~is adequate to ensure no 
straight-line paths longer than 33 \micron, the length of the shortest radon daughter alpha in mylar.  

The DRIFT collaboration installed a prototype texturized thin-film cathode in May of 2013.  
Since then a background rate of only 1 event per day has been observed, 
confirming that texturization can increase the transparency of aluminized mylar to alpha particles.
Efforts toward improving the texturization coverage of the film has continued, 
and new panels with $\geq0.9$ texturized area fraction are ready for installation and further study.

\section{Cathode Rejection Efficiency}
\label{sec:QuantCath}

At this point we have developed the tools, and obtained the data, to determine the efficacy of the various cathodes for vetoing the alpha-decay backgrounds, 
as determined by the figure of merit of Equation \ref{eq:FoM}.  In addition, we are now able to determine the absolute isotopic contamination 
and location of the various sources contributing to the alpha-decays at the cathode. 

To begin, we would like to sub-divide Equation \ref{eq:FoM} into the backgrounds caused by RPRs and those due to LEAs. 
The latter entirely populate the higher energy 
backgrounds from 2000-4000 NIPs, whereas the low energy, 1000-2000 NIPs bin, contains both RPRs and LEAs. 
The background events collected during the runs in question are divided into high and low energy bins, and presented in Table \ref{tab:CleanCont},
along with absolute decay counts of $^{222}$Rn and uranium.

\begin{table}[ht]
 \centering
  \begin{tabular}{l|c|c|c}
    Population & Wire & Dirty & Clean \\
  \hline
  $^{222}$Rn Decays & $64900\pm2200$ & $33800\pm1300$ & $52200\pm1800$ \\
  Uranium Decays & $800\pm53$ & $21900 \pm 500$& $1000\pm50$\\
  \hline
  Backgrounds (low) & $2120\pm46$& $36\pm6$ & $37\pm6$ \\
  Backgrounds (high)& $41\pm6$   &$36\pm6$  & $27\pm5$\\
  \end{tabular}
 \caption{Absolute counts of alpha decays and background counts over various dark matter runs.  
 Low energy backgrounds fall within 1000-2000 NIPs, while high energy events are 2000-4000  NIPS.  
 Uncertainties presented are statistical; systematic uncertainties for absolute alpha counts are all 10\%. 
}
 \label{tab:CleanCont}
\end{table}

These populations can be grouped further.  
High-energy data and a simple model both indicate that the LEA rate is constant over the energy range of interest (1000-4000 NIPS).  
It is therefore assumed that the rate of LEAs in the 1000-2000 NIPS range, where they overlap with RPRs, is the same as the rate in the 2000-4000 NIPS range.  
The total number of LEAs is then given by $150\%$ of the number of high energy backgrounds.  
Similarly, $50\%$ of the high energy backgrounds must be subtracted from the low energy population to remove the contribution by LEAs to obtain the RPR contribution.  

Alpha-producing decays at the cathode can be grouped into those occurring on the surface and those buried in the film.
This grouping is summarized below in Table \ref{tab:SurfaceBuried} in which the neutral fraction of $^{218}$Po in the gas, measured in
Section \ref{sec:neutral}, is denoted by $n=0.23$.  

\begin{table}
 \centering
 \begin{tabular}{l|l|l}
  Isotope & Surface & Buried \\
  \hline
  $^{214}$Po & $\left(\frac{1-n}{2}+n\right)N_{Rn}$ & $\left(\frac{1-n}{2}\right)N_{Rn}$ \\
  $^{218}$Po & $(1-n)N_{Rn}$                        & 0 \\
  U (Both)   & 0                                    & $N_U$ \\
  \hline
  Total      & $1.38N_{Rn}$ & $0.38N_{Rn}$+$N_U$\\
 \end{tabular}
\caption{Counting the number of decays which occur on the surface or buried in the film.  $N_{Rn}$ is the absolute number of $^{222}$Rn decays, and
$N_U$ is the total number of decays from both $^{234}$U and $^{238}$U.  See text for details.}
\label{tab:SurfaceBuried}
\end{table}

After a $^{222}$Rn decay, $n$ of the $^{218}$Po atoms remain neutral in the gas and cannot produce backgrounds.  
The remaining $(1-n)$ of these atoms are charged and electrodeposit onto the central cathode, later decaying on its surface.  
After this decay produces $^{214}$Pb, half of the atoms on the surface ($\frac{1-n}{2}$) bury themselves into the film
while half of them are ejected into the gas as RPRs.  
With a neutral fraction of 0, all of the $^{214}$Pb atoms in the gas, equal to $(\frac{1-n}{2}+n)$ of the total number, 
electrodeposit onto the surface of the cathode.  
Finally, all of the uranium, known to be distributed throughout the aluminum layer (Section \ref{sec:UCont}), is considered to be buried.  
This categorization of the absolute number of alpha decays at the cathode, along with the two types of background (RPR and LEA), 
is presented in Table \ref{tab:Grouped}.

\begin{table}
 \centering
 \begin{tabular}{l|c|c|c}
    Population & Wire & Dirty & Clean \\
    \hline
    Surface Decays& $90000\pm3000$ & $46700\pm1800$ & $72000\pm2500$\\
    Buried Decays & $24700\pm800$  & $34700\pm700$  & $20800\pm700$ \\
    \hline
    RPRs & $2100\pm50$ & $18\pm7$ & $24\pm7$\\
    LEAs & $62\pm10$ & $54\pm9$ & $41\pm8$\\
 \end{tabular}
 \caption{Number of alpha decays compared with background events, grouped into relevant categories.  }
 \label{tab:Grouped}
\end{table}

Table \ref{tab:RejectionEff} presents a comparison of the figure of merit for different cathodes, which leads to several important conclusions.   
The Dirty and Clean thin-film runs, with cathodes identical save for the vastly different contamination levels of buried uranium, 
see LEA/All Alpha production rates which agree to within statistical errors.  
This supports the LEA model of background production which hypothesizes that these backgrounds can be produced by decays 
from atoms deposited at the cathode's surface as well as those found deeper in the material. 
Finally, a comparison of the RPR/Surface decay rate reveals a factor of $70\pm20$ reduction in the probability of producing RPR events when switching from
wire to a thin-film cathode.  This corresponds to the factor of rejection by alpha tag improving from $97.7\%$ to $99.97\%$ between the wire and thin-film runs.  

\begin{table}[ht]
 \centering
 \begin{tabular}{l|c|c|c}
    Population & Wire & Dirty & Clean \\
    \hline
    RPR/Surface & $0.0234\pm0.0007$   & $0.00038\pm0.00014$ & $0.00033\pm0.00009$ \\
    LEA/All     & $0.00052\pm0.00008$ & $0.00044\pm0.00007$ & $0.00042\pm0.00008$ \\
    All/All     & $0.0184\pm0.0006$   & $0.00059\pm0.00007$ & $0.00066\pm0.00008$ \\
 \end{tabular}
 \caption{The probabilities of producing different classes of backgrounds per relevant alpha-producing decay.   }
 \label{tab:RejectionEff}
\end{table}

The background events remaining in DRIFT-IId are now due almost entirely to radon progeny.  
Using the absolute numbers of alpha decays from Table \ref{tab:CleanCont}, the remaining uranium contamination is estimated to produce about $3\%$ of the
LEA backgrounds in DRIFT-IId.\footnote{This fraction is probably slightly higher because the alphas from uranium decays, with shorter tracks, are likely to
have higher background production rates than those from polonium decays.}
Applying the production efficiencies of Table \ref{tab:RejectionEff} to the absolute alpha counts in Table \ref{tab:CleanCont}
indicates that $2\%$ of all backgrounds in DRIFT-IId are now due to uranium contamination of the cathode, 
with the remaining $98\%$ due to RPRs and LEAs produced by radon's polonium daughters.  

\section{Conclusions}

The background events in the DRIFT-IId dark matter detector are due to alpha decays at the central cathode that are partially or fully absorbed by the cathode.  
The wire cathode has been replaced with a thin (0.9 \micron) aluminized mylar cathode to reduce the production rate of these background events by a factor of 
$70\pm20$.  
This new cathode had an unexpected contamination of alpha-producing uranium isotopes, identified using alpha range spectroscopy, which contributes 
further background events.  
The study of these alpha particles pinpoints the uranium location to the thin aluminum coating of the aluminized mylar cathode, 
and measures the contamination to the ppt level.

A new central cathode was built, this time using film produced with uranium-free aluminum.  
This reduced the uranium contamination level by a factor of 10; enough that the remaining uranium is only a minor contributor to the overall background rate.  
The study of alpha particle tracks in DRIFT-IId has increased our understanding of the remaining backgrounds, provided a precise and in-situ assay of the
central cathode, and helped to direct efforts for future improvements.  
These practical results attest to the usefulness of studying these decays in nuclear recoil experiments.  

\section{Acknowledgements}
We acknowledge the support of the US National Science Foundation (NSF). This
material is based upon work supported by the NSF under Grant Nos. 1103420, 1103511, 1407773, and 1506237. 
JBRB acknowledges support from the Sloan Foundation and the Research Corporation.
We also acknowledge support from the Leverhulme Trust through grant PRG-196 at Sheffield.  
We are grateful to Cleveland Potash Ltd and the Science and Technology
Facilities Council (STFC) for operations support and use of the Boulby Underground
Science Facility.

\appendix
\appendixpage

\section{Time Correlation Derivation}
\label{ap:TCorr}
Consider a particular run of duration $T$ while the detector is in equilibrium; the number of observed decays are $N_{Rn}=N \eta_{Rn}$ and $N_{Po}= N \eta_{Po}$.  
  The atoms which decay during this run are indexed by $i\in [1...N]$ such that
for each $i$, $Po_i$ and $Rn_i$ correspond to the times that one atom undergoes two different decays.  The probability distribution for the observation of an individual
decay is flat, corresponding to a constant rate, and is given by Equations \ref{eq:TDistRn} and \ref{eq:TDistPo}.
These are used to calculate the distribution
of differences between all pairs of observed decays, $D(t)$.  Below, $\chi_{[A,B]}(t)$ is a step function whose value is 1 between $A$ and $B$, and $H(t)$ is
the heavyside step function which is 1 for $t\geq0$.  

\begin{equation} 
\label{eq:TDistRn}
P_{Rn_i}(t) =\eta_{Rn}\frac{1}{T}\chi_{[0, T]}
\end{equation}

\begin{equation} 
\label{eq:TDistPo}
P_{Po_i}(t)=\eta_{Po}\frac{1}{T}\chi_{[0, T]}
\end{equation}

\begin{equation}
\label{eq:D_def}
D(t) = \sum_{i=1}^N \sum_{j=1}^N Po_i - Rn_j
\end{equation}

For $D(t)$ there are two distinct cases: when $i\neq j$ and two different atoms are concerned there is no correlation between the timing of these two 
events; when $i=j$ the difference in time is dictated by the decay time $\lambda$ of $^{218}$Po, the isotope which decays second.  

\begin{equation}
\begin{split}
 \label{eq:D_long}
 D(t) & = \sum_{i=1}^N \sum_{j=1}^N \begin{cases}
         i=j, & \eta_{Rn}\eta_{Po}\frac{1}{\lambda}e^{-t/\lambda}H(t) \\
         i\neq j, & \eta_{Rn}\eta_{Po}\frac{T-|t|}{T^2}\chi_{[-T,T]} 
        \end{cases} \\
 D(t) & = N\eta_{Rn}\eta_{Po}\frac{1}{\lambda}e^{-t/\lambda}H(t) + \eta_{Rn}\eta_{Po}\frac{T-|t|}{T^2}(N^2-N)\chi_{[-T,T]} 
\end{split}
\end{equation}

The second term of Equation \ref{eq:D_long} is a background term.  Due to the nature of actual run durations and time-dependent radon concetrations it may 
be more complicated than shown.  Here it is modeled as a second order polynomial, symmetric around 0, and Figure \ref{fig:DecayCorr} shows an exponential 
decay plus this background fit to the distribution of timing differences.   $D'(t)$, used below, is $D(t)$ with this background term removed.  

\begin{equation}
 \label{eq:D_prime}
 D'(t) = N\eta_{Rn}\eta_{Po}\frac{1}{\lambda}e^{-t/\lambda}H(t)
\end{equation}
\begin{equation}
\label{eq:eff}
 \frac{1}{N\eta_{Po}}\int_{-\infty} ^{\infty} \! \mathrm{d}t D'(t) = \frac{N\eta_{Rn}\eta_{Po}}{N\eta_{Po}} = \eta_{Rn} 
\end{equation}

\bibliographystyle{elsarticle-num}
\bibliography{AlphaPaper}

\end{document}